\documentclass[11pt]{article}
\usepackage{graphicx,color,colortbl}
\usepackage{epstopdf}
\usepackage{subfig}
\usepackage{setspace}
\usepackage{natbib}
\usepackage{enumitem}
\usepackage{pgf,tikz}
\usepackage{mathtools}
\usepackage{bm}
\usepackage{multibib} 
\newcites{SM}{References for Online Appendix}
\usepackage{minitoc} 
\usepackage[thin,thinp,thinc]{esdiff}
\usetikzlibrary{positioning}
\usepackage[margin=1in]{geometry}
\definecolor{Gray}{gray}{0.85} 
\newcolumntype{a}{>{\columncolor{Gray}}c} 
\definecolor{mycol}{rgb}{0.64, 0.0, 0.0}
\definecolor{amber}{rgb}{1.0, 0.49, 0.0}
\definecolor{navyblue}{rgb}{0.0, 0.0, 0.5}
\definecolor{firebrick}{rgb}{0.7, 0.13, 0.13}
\newcommand{\indep}{\mathop{\perp\!\!\!\!\perp}}
\newcommand\numberthis{\addtocounter{equation}{1}\tag{\theequation}}
\usepackage{etoolbox}
\makeatletter

\patchcmd{\NAT@citex}
  {\@citea\NAT@hyper@{%
     \NAT@nmfmt{\NAT@nm}%
\hyper@natlinkbreak{\NAT@aysep\NAT@spacechar}{\@citeb\@extra@b@citeb}%
     \NAT@date}}
  {\@citea\NAT@nmfmt{\NAT@nm}%
   \NAT@aysep\NAT@spacechar\NAT@hyper@{\NAT@date}}{}{}

\patchcmd{\NAT@citex}
  {\@citea\NAT@hyper@{%
     \NAT@nmfmt{\NAT@nm}%
\hyper@natlinkbreak{\NAT@spacechar\NAT@@open\if*#1*\else#1\NAT@spacechar\fi}%
       {\@citeb\@extra@b@citeb}%
     \NAT@date}}
  {\@citea\NAT@nmfmt{\NAT@nm}%
\NAT@spacechar\NAT@@open\if*#1*\else#1\NAT@spacechar\fi\NAT@hyper@{\NAT@date}}
  {}{}

\makeatother 

\usepackage[colorlinks=true,
            linkcolor=mycol,
            citecolor=mycol,
            ]{hyperref}

\usepackage{rotating}
\usepackage{pdfpages}
\usepackage{dcolumn}
\usepackage{amssymb}
\usepackage{amsmath}
\usepackage{multicol}
\usepackage{multirow}
\usepackage{dsfont}
\newcommand{\E}{\mathbb{E}}
\newcommand{\V}{\mathbb{V}}

\usepackage{setspace}
 \begin{titlepage}
 \title{A Bias-Corrected Estimator for the Crosswise Model with Inattentive Respondents\thanks{An open-source software \textsf{R} package \textsf{cWise:}\textsf{A (Cross)Wise Method to Analyze Sensitive Survey Questions}, which implements our methods, is available at \href{https://github.com/YukiAtsusaka/cWise}{\color{mycol} \textsf{https://github.com/YukiAtsusaka/cWise}}. The previous version of this manuscript was titled as ``Bias-Corrected Crosswise Estimators for sensitive questions.'' For helpful comments and valuable feedback, we would like to thank Graeme Blair, Gustavo Guajardo, Dongzhou Huang, Colin Jones, Gary King, Shiro Kuriwaki, Jeff Lewis, John Londregan, Yui Nishimura, Michelle Torres, and three anonymous reviewers.}}
 \author{\Large Yuki Atsusaka\thanks{Ph.D. Candidate, Department of Political Science, Rice University, MS-24 105 Herzstein Hall, 6100 Main St, Houston, TX 77005 (\href{https://atsusaka.wordpress.com/}{\color{mycol} \textsf{https://atsusaka.org}}).}\vspace{0.1cm} \\  \href{mailto:atsusaka@rice.edu}{\textsf{\Large \color{mycol}atsusaka@rice.edu}}
  \and \Large Randolph T. Stevenson\thanks{Professor of Political Science, Rice University, MS-24 105 Herzstein Hall, 6100 Main St, Houston, TX 77005.}\vspace{0.1cm}  \\ \href{mailto:randystevenson@rice.edu}{\textsf{\Large \color{mycol}randystevenson@rice.edu}}
}
 \date{\vspace{0.3cm} \today}
 \end{titlepage}
\begin{document}
\doparttoc 
\faketableofcontents 
\maketitle
\thispagestyle{empty} 
\begin{abstract}
The crosswise model is an increasingly popular survey technique to elicit candid answers from respondents on sensitive questions. Recent studies, however, point out that in the presence of inattentive respondents, the conventional estimator of the prevalence of a sensitive attribute is biased toward 0.5. To remedy this problem, we propose a simple design-based bias correction using an anchor question that has a sensitive item with known prevalence. We demonstrate that we can easily estimate and correct for the bias arising from inattentive respondents without measuring individual-level attentiveness. We also offer several useful extensions of our estimator, including a sensitivity analysis for the conventional estimator, a strategy for weighting, a framework for multivariate regressions in which a latent sensitive trait is used as an outcome or a predictor, and tools for power analysis and parameter selection. Our method can be easily implemented through our open-source software, \textsf{cWise}.\\

\end{abstract} 

\indent

\noindent \textsf{Word Count: 10135}


\newpage
\setcounter{page}{1} 
\doublespacing
\section{Introduction}
Political scientists often use surveys to estimate and analyze the prevalence of sensitive attitudes and behaviors.\footnote{Such topics include racial animus \citep{kuklinski1997racial}, attitudes toward non-Christian candidates \citep{kane2004religion}, support for militant organizations \citep{lyall2013explaining}, support for authoritarian regimes \citep{frye2017putin}, voting for a right-wing populist party \citep{lehrer2019wisdom}, and vote buying \citep{cruz2019social}.} To mitigate ``sensitivity bias'' in self-reported data \citep{blair2020worry}, such as bias arising from social desirability, self-image protection, fear of disclosing truth, and perceived intrusiveness, various survey techniques have been developed including randomized response techniques, list experiments, and endorsement experiments \citep{rosenfeld2016empirical}. The \textit{crosswise model} is an increasingly popular design among these techniques.\footnote{About 60 studies related to the design were published between 2016 and 2021 across disciplines.} A crosswise question shows respondents two statements: one about a sensitive attitude or behavior of interest (e.g., ``I am willing to bribe a police officer'') and one about some piece of private, non-sensitive information whose population frequency is known (e.g., ``My mother was born in January''). The respondents are then asked a question, the answer to which depends jointly on the truth status of both statements and so fully protects respondent privacy \citep{yu2008two}. The key idea to the design is that even though researchers only observe respondents' answers to the joint condition, they can estimate the population prevalence of the sensitive attribute using the known probability distribution of the non-sensitive statement.\footnote{This idea of intentionally injecting statistically tractable noise originated from the randomized response technique, which was inherited by the literature on differential privacy \citep{evans2021statistically}.}

Despite its promise and several advantages over other indirect questioning techniques \citep[see e.g.,][]{meisters2020controlling}, recent studies suggest that the crosswise model suffers from two interrelated problems, casting doubt on the validity of the design. First, in many settings, its relatively complex format (compared to direct questioning) leads to a significant number of inattentive survey respondents who give answers that are essentially random \citep{enzmann2017,john2018and,heck2018detecting,schnapp2019sensitive,walzenbach2019pouring}.\footnote{One strength of the design is that there is not a clear best strategy for non-cooperators and so by including some simple design features (discussed below), most non-cooperative response strategies can be made to be ``as if'' random. As such, we refer to any such strategy as random or ``inattentive.''} Second, this tendency usually results in overestimates of the prevalence of sensitive attributes \citep{hoglinger2017uncovering,hoglinger2018more,nasirian2018does,meisters2020can}. While several potential solutions to this problem have been discussed in the extant literature \citep{enzmann2017,schnapp2019sensitive,meisters2020can}, they rely on rather strong assumptions and external information from either attention checks or a different unprotected --- but still sensitive --- question answered by the same respondents, leading to unsatisfying solutions.\footnote{Several other studies have contributed important work relevant to building a solution, though each stops short of actually offering one \citep{hoglinger2017uncovering,hoglinger2018more,walzenbach2019pouring,meisters2020can}.} In this article, we propose an alternative solution that builds on insights about ``zero-prevalence items'' \citep{hoglinger2017uncovering} to correct for the bias arising from inattentive respondents.


More generally, we provide the first detailed description and statistical evaluation of methods for measuring and mitigating the bias caused by inattentive respondents in the crosswise model. This includes an evaluation of the performance of our new estimator and a brief assessment of previous methods. Unlike previous efforts, this allows us to not only offer our method as a solution for estimating prevalence rates without bias, but also to explain how its assumptions can be easily evaluated and made to hold by design. It also allows us to develop a number of extensions that enhance its practical usefulness to researchers. These include (1) a sensitivity analysis to simulate the amount of bias caused by inattentive respondents even when our correction is not available; (2) a weighting strategy that allows the estimator to be used with general sampling schemes; (3) a framework for multivariate regressions in which a latent sensitive trait is used as an outcome or a predictor; and (4) simulation-based tools for power analysis and parameter selection. 


In what follows, we first describe the crosswise model and demonstrate that the conventional estimator is biased toward 0.5 when inattentive respondents are present (Section \ref{sec:crosswise}). Next, we introduce a simple design-based solution that enables researchers to estimate and correct for this bias without measuring \textit{individual-level} attentiveness for each respondent. Specifically, we propose including a certain kind of \textit{anchor question} in the survey while making several assumptions whose plausibility can be enhanced with simple modifications to the survey design (Section \ref{sec:correction}). Next, we evaluate the proposed estimator in a series of simulations that demonstrate our bias-corrected estimator can recover the ground truth (Section \ref{sec:simulation}). The final sections of the paper present the five extensions of our bias-corrected estimator (Section \ref{sec:extension}) and concludes with practical suggestions (Section \ref{sec:guide}). We offer easy-to-use software that allows users to analyze data from the crosswise model with our bias correction and extensions.


\section{Promise and Pitfalls of the Crosswise Model}\label{sec:crosswise}

\subsection{The Crosswise Model}
The crosswise model was developed by \citet{yu2008two} to overcome several limitations with randomized response techniques \citep{warner1965randomized,blair2015design}. In political science and related disciplines, the design has been used to study corruption \citep{corbacho2016corruption,oliveros2020lying}, ``shy voters'' and strategic voting \citep{waubert2017indirect}, self-reported turnout \citep{kuhn2018reducing}, prejudice against female leaders \citep{hoffmann2019prejudice}, xenophobia \citep{hoffmann2016assessing}, and anti-refugee attitudes \citep{hoffmann2020validity}. The crosswise model asks the respondent to read two statements whose veracity is known only to her. For example, \citet{corbacho2016corruption} study corruption in Costa Rica with the following question:

\begin{figure}[h!]
\centering
\sf 
\fbox{\begin{minipage}{39em}
\small
\indent \textbf{Crosswise Question: How many of the following statements are true?} \\

\indent \quad  \textsf{Statement A: In order to avoid paying a traffic ticket, I would be willing to pay a bribe to a police officer}\\
\indent \quad  \textsf{Statement B: My mother was born in October, November, or December}\\

\indent \quad  $\bullet$ \underbar{\text{Both}} statements are true, or \underbar{\text{neither}} statement is true\\
\indent \quad  $\bullet$ \underbar{\text{One}} of the two statements is true
\end{minipage}}
\end{figure}

\textsf{Statement A} is the sensitive statement (e.g., corruption) that researchers would have asked directly if there had been no worry about sensitivity bias. The quantity of interest is the population proportion of individuals who agree with \textsf{Statement A}.\footnote{Often, this proportion is less than 0.5 and direct questioning is expected to underestimate the quantity. Thus, we assume that the quantity of interest is always less than 0.5 in the rest of the argument. Even when it is greater than 0.5 (e.g., most people say Yes to the statement), we can always word the sensitive statement to flip the expected direction (i.e., including ``NOT'' clause) or adjust estimates post-survey (i.e., subtracting the quantity of interest from 1).} In contrast, \textsf{Statement B} is a non-sensitive statement (e.g., mother's birth month) whose population prevalence is \textit{ex ante} known to researchers.\footnote{Prevalence rates for non-sensitive questions can come from census data or other kinds of statistical regularities like the Newcomb-Benford law \citep{kundt2014applying}. Further, in Section \ref{app:secret}, we present an approach that relies on a virtual die roll but that overcomes the respondent's natural skepticism that such information will be recorded.} The crosswise model then asks respondents whether ``both statements are true, or neither statement is true'' or ``one of the two statements is true.'' Respondents may also have the option of choosing ``Refuse to Answer'' or ``Don't Know.'' 



Importantly, the respondent's answer does not allow interviewers (or anyone) to know whether the individual agrees or disagrees with the sensitive statement, which fully protects the respondent's privacy. Nevertheless, the crosswise model allows us to compute the proportion of respondents for which the sensitive statement is true via a simple calculation. Suppose that the observed proportion of respondents choosing ``both or neither is true'' is 0.65 while the known population proportion for \textsf{Statement B} is 0.25. If the sensitive and non-sensitive statements are statistically independent, it follows that:
$\widehat{\mathbb{P}}(\textsf{TRUE-TRUE} \cup \textsf{FALSE-FALSE}) = 0.65 \Rightarrow  \widehat{\mathbb{P}}(\textsf{A=TRUE})\times\mathbb{P}(\textsf{B=TRUE}) +\widehat{\mathbb{P}}(\textsf{A=FALSE})\times\mathbb{P}(\textsf{B=FALSE}) = 0.65
\Rightarrow \widehat{\mathbb{P}}(\textsf{A=TRUE})\times0.25 + ( 1 - \widehat{\mathbb{P}}(\textsf{A=TRUE}))\times 0.75 = 0.65
\Rightarrow \widehat{\mathbb{P}}(\textsf{A=TRUE}) = \frac{0.65-0.75}{-0.5}=0.2.
$, where $\widehat{\mathbb{P}}$ is an estimated proportion.



\subsection{Relative Advantages and Limitations}
Despite its increasing popularity and introduction by \citet{corbacho2016corruption} and \citet{gingerich2016protect}, the crosswise model has not been widely used in political science. We think that the primary reason is that it has not been clear to many political scientists how and when the crosswise model is preferable to other indirect questioning techniques. To help remedy this problem, Table \ref{tab:Advantages} summarizes the relative advantages of the crosswise model over randomized response, list experiments, and endorsement experiments, which are more commonly used survey techniques in political science \citep[for more detailed comparisons and discussions, see e.g.,][]{blair2014comparing,hoffmann2015strong,rosenfeld2016empirical,hoglinger2018more,blair2020worry,meisters2020controlling}.\footnote{We only consider the simplest (benchmark) design for each technique.} 

\begin{table}[h!]
\centering
\small
\begin{tabular}{lacccc} \hline
                             &  \textbf{Crosswise} & RR & List & Endorsement  \\ 
\hline
Privacy protection & full protection & full protection & partial to full protection &  full protection &  \\
Self-protective option    & no & yes  & yes & yes  \\ 
Split samples     & not required    & not required & required  &  required & \\
Randomization device     & not required    & required & not required  & not required & \\ 
Efficiency           & high & high & moderate & low\\
Measurement         & direct & direct  & direct & indirect \\
Instruction complexity       & moderate        & difficult & easy  & very easy & \\ 
Auxiliary data       & required & required  & not required & not required \\
   \hline
\end{tabular}
\caption{ \textbf{Relative Advantages of the Crosswise Model.} \textit{Note}: This table shows potential (dis)advantages of the crosswise model compared to randomized response techniques (RR), list experiments (List), and endorsement experiments (Endorsement).}\label{tab:Advantages}
\end{table}

The potential advantages of the crosswise model are that the design (1) fully protects respondents' privacy,\footnote{The crosswise model does not have any ceiling and floor effects that reveal respondents' true answer as in list experiments. Recent studies, however, show that the \textit{perceived} level of privacy protection given by the crosswise model may not be as high as its objective level of confidentiality \citep{hoffmann2017comprehensibility,quatember2019discussion,jerke2019too,meisters2020can}, which suggests more effort in designing survey instructions and simple explanations of the technique is warranted. In our experience, once a respondent understands she is protected, it is also helpful to provide a positive reason to answer candidly.} (2) provides no incentive for respondents to lie about their answers because there is no ``safe'' (self-protective) option, (3) does not require splitting the sample (as in list and endorsement experiments), (4) does not need an external randomization device (as in randomized response), (5) is relatively efficient compared to other designs, and (6) asks about sensitive attributes directly (unlike in endorse experiments). In contrast, the potential disadvantage of this design is that its instruction may be harder to understand compared to those of list and endorsement experiments (but are likely much easier than for randomized response). Additionally, the crosswise model requires auxiliary data in the form of a known population distribution of the non-sensitive information (as in randomized response). 



A validation study by \citet{rosenfeld2016empirical} shows that randomized response appears to outperform list and endorsement experiments by yielding the least biased and most efficient estimate of the ground truth. Since the crosswise model was developed to outperform randomized response \citep{yu2008two}, the crosswise model is expected to better elicit candid answers from survey respondents. To date, several validation studies appear to confirm this expectation \citep{jann2011asking,hoffmann2015strong,hoffmann2016assessing,hoglinger2016sensitive,hoglinger2018more,meisters2020can,jerke2020handle}.

Recently, however, it has been increasingly clear that the design has two major limitations, which may undermine confidence in the method. First, it produces a relatively large share of inattentive respondents who give random answers \citep{enzmann2017,john2018and,heck2018detecting,schnapp2019sensitive}. Observing this tendency,  \citet[14]{walzenbach2019pouring} even concludes that ``a considerable number of respondents do not comply with the intended procedure''  and it ``seriously limit[s] the positive reception the [crosswise model] has received in the survey research so far.'' Second, it appears to overestimate the prevalence of sensitive attributes and yields relatively high \textit{false positive} rates \citep{hoglinger2018more,nasirian2018does,kuhn2018reducing,meisters2020can}. After finding this ``blind spot,'' \citet[135]{hoglinger2017uncovering} lament that ``[p]revious validation studies appraised the [crosswise model] for its easy applicability and seemingly more valid results. However, none of them considered false positives. Our results strongly suggest that in reality the [crosswise model] as implemented in those studies does not produce more valid data than [direct questioning].'' Nevertheless, we argue that with a proper understanding of these problems, it is possible to solve them and even extend the usefulness of the crosswise model. To do so, we need to first understand how inattentive respondents lead to bias in estimated prevalence rates.


\subsection{Inattentive Respondents under the Crosswise Model}\label{sec:inattentive}
The problem of inattentive respondents is well-known to survey researchers, who have used a variety of strategies for detecting them such as attention checks  \citep{oppenheimer2009instructional}. Particularly in self-administered surveys, estimates of the prevalence of inattentiveness are often as high as 30 to 50\% \citep{berinsky2014separating}. Recently, inattention has also been discussed with respect to list experiments \citep{ahlquist2018list,blair2019list}. Inattentive respondents are also common under the crosswise model, as we might expect given its relatively complex instructions \citep{alvarez2019paying}. Researchers have estimated the proportion of inattentive respondents in surveys featuring the crosswise model to be from 12\% \citep{hoglinger2017uncovering} to 30\% \citep{walzenbach2019pouring}.\footnote{The crosswise model has been implemented in more than 8 countries with different platforms and formats and inattention has been a concern in many of these studies.} 







To proceed, we first define inattentive respondents under the crosswise model as \textit{respondents who randomly choose between ``both or neither is true'' and ``only one statement is true.''} We assume that random answers may arise due to multiple reasons, including \textit{nonresponse}, \textit{noncompliance}, \textit{cognitive difficulty}, \textit{lying}, or any combination of these \citep{heck2018detecting,jerke2019too,meisters2020can}. 

\begin{figure}[tbh!]
    \centering
    \includegraphics[width=12cm]{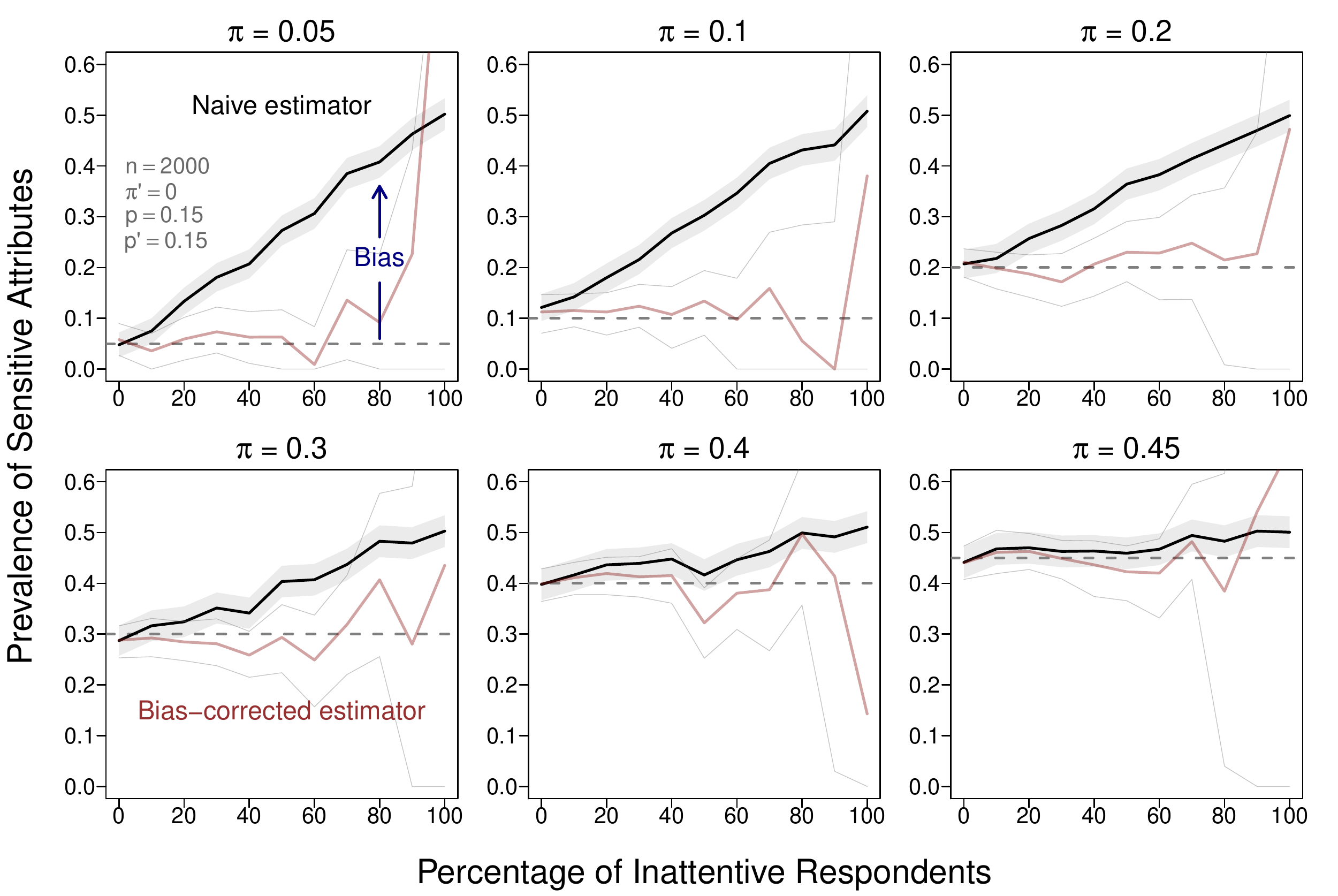}
    \caption{ \textbf{Consequences of Inattentive Respondents.} \textit{Note}: This figure illustrates how the conventional estimator (solid line) is biased toward 0.5, whereas the proposed bias-corrected estimator captures the ground truth (dashed horizontal line). The bias increases as the percentage of inattentive respondents increases and the true prevalence rate ($\pi$) decreases. The top-left panel notes parameter values for all six panels (for notation, see the next section).} 
\label{fig:simulate0}
\end{figure}

We now consider the consequences of inattentive respondents in the crosswise model.\footnote{Of course, this will not be a problem if researchers define the prevalence of sensitive attributes \textit{only among} attentive respondents as their quantity of interest.} Figure \ref{fig:simulate0} plots the bias in the conventional estimator based on hypothetical (and yet typical) data from the crosswise model.\footnote{Several studies have conjectured that the presence of inattentive respondents biases the point estimate toward 0.5, such as Appendix 7 of \citet{john2018and}, Figure C.4 in the Online Appendix of \citet{hoglinger2017uncovering}, \citet[1899]{heck2018detecting}, \citet{enzmann2017}, and \citet[311]{schnapp2019sensitive}).} The figure clearly shows the expected bias toward 0.5 and suggests that the bias grows as the percentage of inattentive respondents increases and as the quantity of interest (labeled as $\pi$) gets close to 0. We also find that the size of bias does not depend on the prevalence of the non-sensitive item. To preview the performance of our bias-corrected estimator we introduce below, each plot also shows estimates based on that estimator, which is robust to the presence of inattentive respondents regardless of the value of the quantity of interest. One key takeaway is that researchers must be more cautious about inattentive respondents when the quantity of interest is expected to be close to zero because these cases tend to produce larger biases.

Figure \ref{fig:simulate0} also speaks to critiques of the crosswise model that have focused on the incidence of false positives, as opposed to bias overall \citep{hoglinger2017uncovering,hoglinger2018more,nasirian2018does,meisters2020can}. While these studies are often agnostic about the source of false-positives, the size of the biases in this figure suggests that the main source is likely inattentive respondents.


Several potential solutions to this problem have been recently discussed. The first approach is to identify inattentive survey takers via comprehension checks, remove them from data, and perform estimation and inference on the ``cleaned'' data (i.e., listwise deletion) \citep{hoglinger2017uncovering,hoglinger2018more,meisters2020can}. One drawback of this method is that it is often challenging to discover which respondents are being inattentive. Moreover, this approach leads to a biased estimate of the quantity of interest unless researchers make the \textit{ignorability} assumption that having a sensitive attribute is statistically independent of one's attentiveness \citep[148]{alvarez2019paying}, which is a reasonably strong assumption in most situations.\footnote{For a more general treatment of the consequences of dropping inattentive respondents in randomized experiments, see \citet{aronow2019note}.} The second solution detects whether respondents answered the crosswise question randomly via direct questioning and then adjusts the prevalence estimates accordingly \citep{enzmann2017,schnapp2019sensitive}. This approach is valid if researchers assume that direct questioning is itself not susceptible to inattentiveness or social desirability bias as well as that the crosswise question does not affect respondents' answers to the direct question. Such an assumption, however, is highly questionable and the proposed corrections have several undesirable properties (Section \ref{sec:appSimulation}). Below, we present an alternative solution to the problem, which yields an unbiased estimate of the quantity of interest with a much weaker set of assumptions than in existing solutions.

\section{The Proposed Methodology}\label{sec:correction}
\setcounter{section}{3}


\subsection{The Setup}
Consider a single sensitive question in a survey with $n$ respondents drawn from a finite target population via simple random sampling with replacement. Suppose that there are no missing data and no respondent opts out of the crosswise question.\footnote{If some respondents opt-out, researchers may utilize our proposed weighting strategy (Section \ref{sec:weighting}).} Let $\pi$ (quantity of interest) be the population proportion of individuals who agree with the sensitive statement (e.g., I am willing to bribe a police officer). Let $p$ be the known population proportion of people who agree with the non-sensitive statement (e.g., My mother was born in January). Finally, let $\lambda$ (and 1 - $\lambda$) be the population proportion of individuals who choose ``both or neither is true'' (and ``only one statement is true''). Assuming $\pi \indep p$, \citet{yu2008two} introduced the following identity as a foundation of the crosswise model:
\begin{subequations}
\begin{equation}
\mathbb{P}(\text{TRUE-TRUE} \cup \text{FALSE-FALSE}) = \lambda = \pi p + (1 - \pi)(1-p).\label{eq:crosswise}
\end{equation}


\noindent Solving the identity with respect to $\pi$ yields $\pi = \frac{\lambda + p - 1}{2p - 1}$. Based on this identity, the authors proposed the na\"ive crosswise estimator:
\begin{equation}
\widehat\pi_{CM} = \frac{\widehat\lambda + p - 1}{2p - 1},\label{eq:naive}
\end{equation}

\noindent where $\widehat \lambda$ is the observed proportion of respondents choosing ``both or neither is true'' and $p \neq 0.5$.

We call Equation (\ref{eq:naive}) the na\"ive estimator because it does not take into account the presence of inattentive respondents who give random answers in this design. When one or more respondents do not follow the instruction and randomly pick their answers, the proportion must be \citep[generalizing][10]{walzenbach2019pouring}:
\begin{equation}
\lambda = \Big\{ \pi p + (1-\pi)(1-p) \Big\}\gamma + \kappa(1-\gamma)\label{eq:lambda},
\end{equation}  

\noindent where $\gamma$ is the proportion of \textit{attentive} respondents and $\kappa$ is the probability with which inattentive respondents pick ``both or neither is true.''\footnote{Note that Equation (\ref{eq:lambda}) is a strict generalization of Equation (\ref{eq:crosswise}). While the na\"ive estimator is unbiased as long as $\gamma=1$ (Section \ref{sec:naive}), it is no longer an unbiased estimator whenever $\gamma < 1$.}

We then quantify the bias in the na\"ive estimator as follows (Section \ref{sec:bias}):
\begin{align}
 B_{CM} & \equiv \E[\widehat\pi_{CM}] - \pi \\
  & = \Bigg( \frac{1}{2} - \frac{1}{2\gamma} \Bigg)\Bigg( \frac{\lambda - \kappa}{p-\frac{1}{2}} \Bigg).\label{eq:Bias}
\end{align}
\end{subequations}  

Here, $B_{CM}$ is a bias with respect to our quantity of interest caused by inattentive respondents. Under regularity conditions ($\pi < 0.5$, $p < \frac{1}{2}$, and $\lambda > \kappa$), which are met in typical crosswise models, the bias term is always positive (Section \ref{sec:appBias}). This means that the conventional crosswise estimator always \textit{overestimates} the population prevalence of sensitive attributes in the presence of inattentive respondents. 

\subsection{Bias-Corrected Crosswise Estimator}
To address this pervasive issue, we propose the following bias-corrected crosswise estimator:
\begin{subequations}
\begin{align}
\widehat \pi_{BC} =\widehat\pi_{CM} - \widehat B_{CM},\label{eq:piBC}
\end{align}
\noindent where $\widehat B_{CM}$ is an unbiased estimator of the bias:
\begin{align}
\widehat B_{CM} = \Bigg( \frac{1}{2} - \frac{1}{2\widehat\gamma} \Bigg)\Bigg( \frac{\widehat\lambda - \frac{1}{2}}{p-\frac{1}{2}} \Bigg)\label{eq:biashat}
\end{align}
\end{subequations}

\noindent and $\widehat{\gamma}$ is the estimated proportion of attentive respondents in the crosswise question (we discuss how to obtain $\widehat{\gamma}$ below).\footnote{Following \citet{yu2008two}, we further impose a logical constrain:$\widehat \pi_{BC} = \min(1, \max(0,\widehat\pi_{BC})).$ As shown below, $\E[\widehat \lambda] = \lambda$ and $\E[\widehat\gamma] = \gamma$ and thus, by the linearity of the expected value operator, $\E[\widehat B_{CM}] = B_{CM}$.} 

This bias correction depends on several assumptions. First, we assume that inattentive respondents choose ``both or neither is true'' with probability 0.5.\\

\noindent \textbf{Assumption 1 (Random Pick).} \textit{Inattentive respondents choose ``both or neither is true'' with probability 0.5 (i.e., $\kappa=0.5$).}\\

Though many studies appear to take Assumption 1 for granted, the survey literature suggests that this assumption may not hold in many situations because inattentive respondents tend to choose first listed items more than second (or lower) listed ones \citep{krosnick1991response,galesic2008eye}. Nevertheless, it is still possible to \textit{design} a survey to achieve $\kappa=0.5$ regardless of how inattentive respondents choose items. For example, we can achieve this goal by randomizing the order of the listed items in the crosswise model. 


The main challenge in estimating the bias is to obtain the estimated proportion of attentive respondents in the crosswise question (i.e., $\widehat\gamma$). We solve this problem by adding an \textit{anchor question} to the survey. The anchor question uses the same format as the crosswise question, but contains a sensitive statement with \textit{known} prevalence. Our proposed solution generalizes and extends the idea of ``zero-prevalence sensitive items'' first introduced by \citet{hoglinger2017uncovering}.\footnote{More generally, previous research has used sensitive items with known prevalence only for validation purposes. For example, \citet{hoglinger2017uncovering} use receiving an organ and the history of having a rare disease as zero-prevalence sensitive items in the survey about ``organ donation and health.'' Similarly, \citet{rosenfeld2016empirical} employ official county-level results of an anti-abortion referendum, while \citet{kuhn2018reducing} rely on official turnout records as sensitive items with know prevalence.} The essence of our approach is to use this additional sensitive statement to (1) estimate the proportion of inattentive respondents in the anchor question and (2) use it to correct for the bias in the crosswise question. For our running example (corruption in Costa Rica), we might consider the following anchor question:\footnote{Section \ref{sec:appAnchor} introduces multiple examples of sensitive-statements with known prevalence.}


\begin{figure}[h!]
\centering
\sf 
\small
\fbox{\begin{minipage}{38em}
\indent \textbf{Anchor Question: How many of the following statements is true?} \\

\indent \quad  \textsf{Statement C: I have paid a bribe to be on the top of a waiting list for an organ transplant}\\
\indent \quad  \textsf{Statement D: My best friend was born in January, February, or March}\\

\indent \quad  $\bullet$ \underbar{\text{Both}} statements are TRUE, or \underbar{\text{neither}} statement is TRUE \\
\indent \quad  $\bullet$ \underbar{\text{One}} of the two statements is TRUE
\end{minipage}}
    \label{fig:anchorQ}
\end{figure}

Here, \textsf{Statement C} is a sensitive anchor statement whose prevalence is (expected to) be 0.\footnote{More generally, the anchor question can feature any sensitive item whose true prevalence is known (i.e., it need not be 0), though zero-prevalence sensitive items lead to more efficient estimates and may be easier to construct (see Section \ref{sec:appAnchor}).} \textsf{Statement D} is a non-sensitive statement whose population prevalence is known to researchers just like the non-sensitive statement in the crosswise question. Additionally, ``Refuse to Answer'' or ``Don't Know'' may be included.


Let $\pi'$ be the \textit{known} proportion for \textsf{Statement C} and $p'$ be the known proportion for \textsf{Statement D}  (the ``prime symbol'' indicates the anchor question). Let $\lambda'$ (and 1 - $\lambda'$) be the population proportion of people selecting ``both or neither is true'' (and ``only one statement is true'').  Let $\gamma'$ be the population proportion of attentive respondents in the anchor question. The population proportion of respondents choosing ``both or neither is true'' in the anchor question then becomes:
\begin{subequations}
\begin{equation}
\lambda' = \Big\{ \pi' p' + (1-\pi')(1-p') \Big\}\gamma' + \kappa(1-\gamma')\label{eq:lambda.prime}
\end{equation}
    
\noindent Assuming $\kappa=0.5$ (Assumption 1) and $\pi'=0$ (zero-prevalence), we can rearrange Equation (\ref{eq:lambda.prime}) as:
\begin{equation}
\gamma' = \frac{\lambda' - \frac{1}{2}}{\pi'p'-(1-\pi')(1-p')-\frac{1}{2}} = \frac{\lambda' - \frac{1}{2}}{\frac{1}{2}-p'}\label{eq:gamma}.
\end{equation}

\noindent We can then estimate the proportion of attentive respondents \textit{in the anchor question} as:

\begin{equation}
\widehat\gamma' = \frac{\widehat\lambda' - \frac{1}{2}}{\frac{1}{2}-p'}\label{eq:gammaHat},
\end{equation}

\noindent where $\widehat{\lambda'}$ is the observed proportion of ``both or neither is true'' and $\E[\widehat\lambda']=\lambda'$ (Section \ref{sec:lambdaprime}).

Finally, our strategy is to use $\widehat\gamma'$ (obtained from the anchor question) as an estimate of $\gamma$ (the proportion of attentive respondents in the crosswise question) and plug it into Equation (\ref{eq:biashat}) to estimate the bias. This final step yields the complete form of our bias-corrected estimator:
\begin{equation}
\widehat \pi_{BC} =\widehat\pi_{CM} - \underbrace{\Bigg( \frac{1}{2} - \frac{1}{2}\Bigg[\frac{\frac{1}{2}-p'}{\widehat{\lambda}' - \frac{1}{2}}\Bigg] \Bigg)\Bigg( \frac{\widehat\lambda - \frac{1}{2}}{p-\frac{1}{2}} \Bigg) }_{\text{Estimated Bias: } \widehat{B}_{CM} (\widehat{\lambda}, \widehat{\lambda}', p, p') },\label{eq:piBCFULL}
\end{equation}
\end{subequations}

\noindent where it is clear that our estimator depends both on the crosswise question ($\widehat{\lambda}$ and $p$) and the anchor question ($\widehat{\lambda}'$ and $p'$). For the proposed estimator to be unbiased, we need to make two assumptions (Section \ref{sec:proofunbiased}):\\

\noindent \textbf{Assumption 2 (Attention Consistency).} \textit{The proportion of attentive respondents does not change across the crosswise and anchor questions (i.e., $\gamma = \gamma'$).}\\

\noindent \textbf{Assumption 3 (No Carryover).} \textit{The crosswise question does not affect respondents' answers to the anchor question and vice versa (and thus $\E[\widehat{\lambda}]=\lambda$ and $\E[\widehat{\lambda}']=\lambda'$).}\\

Assumption 2 will be violated, for example, if the crosswise question has a higher level of inattention than the anchor question. Assumption 3 will be violated, for instance, if asking the anchor question makes some respondents more willing to bribe a police officer in our running example.

Importantly, researchers can design their surveys to make these assumptions more plausible. For example, they can do so by randomizing the order of the anchor and non-anchor crosswise questions, making them look alike, and using a statement for the anchor question that addresses the same topic and is equally sensitive as the one in the crosswise question. These considerations also help to satisfy Assumption 3 (we provide more practical advice on how to do this in the conclusion and in Section \ref{sec:appAnchor}).

Finally, we derive the variance of the bias-corrected crosswise estimator and its sample analog as follows (Section \ref{sec:variance}):
\begin{subequations}
\begin{align}
\V(\widehat \pi_{BC}) =  \V  \Bigg[\frac{\widehat{\lambda}}{\widehat{\lambda}'}   \Bigg(\frac{\frac{1}{2}-p'}{2p-1} \Bigg) \Bigg] \quad \text{and} \quad
\widehat{\V}(\widehat \pi_{BC}) =  \widehat{\V}  \Bigg[\frac{\widehat{\lambda}}{\widehat{\lambda}'} \Bigg(\frac{\frac{1}{2}-p'}{2p-1} \Bigg) \Bigg]\label{eq:varBChat}
\end{align}
\end{subequations}

Note that these variances are necessarily larger than those of the conventional estimator as the bias-corrected estimator also needs to estimate the proportion of (in)attentive respondents from data. Since no closed-form solutions to these variances are available, we employ the bootstrap to construct confidence intervals.\footnote{Importantly, these equations in (\ref{eq:varBChat}) also explain why choosing a zero-prevalence item for the anchor question is more efficient than selecting a non-zero prevalence item (i.e., when $\pi'=0$, the variability of $\widehat{\lambda}'$ only comes from $p'$ and $\widehat{\gamma}'$).}

In sum, our bias-corrected estimator provides an unbiased estimate of the population prevalence of a sensitive attribute even when some respondents answer randomly. We do so by using only three, rather weak, assumptions that can be easily satisfied in the design stage of the survey. Moreover, our solution avoids a common trade-off between running the risk of having biased estimates (by including inattentive respondents) and inducing selection bias and lower statistical power (by dropping inattentive respondents) \citep[740]{berinsky2014separating}.

\section{Simulation Studies}\label{sec:simulation}
\begin{figure}[tbh!]
\centering
 \includegraphics[width=13cm]{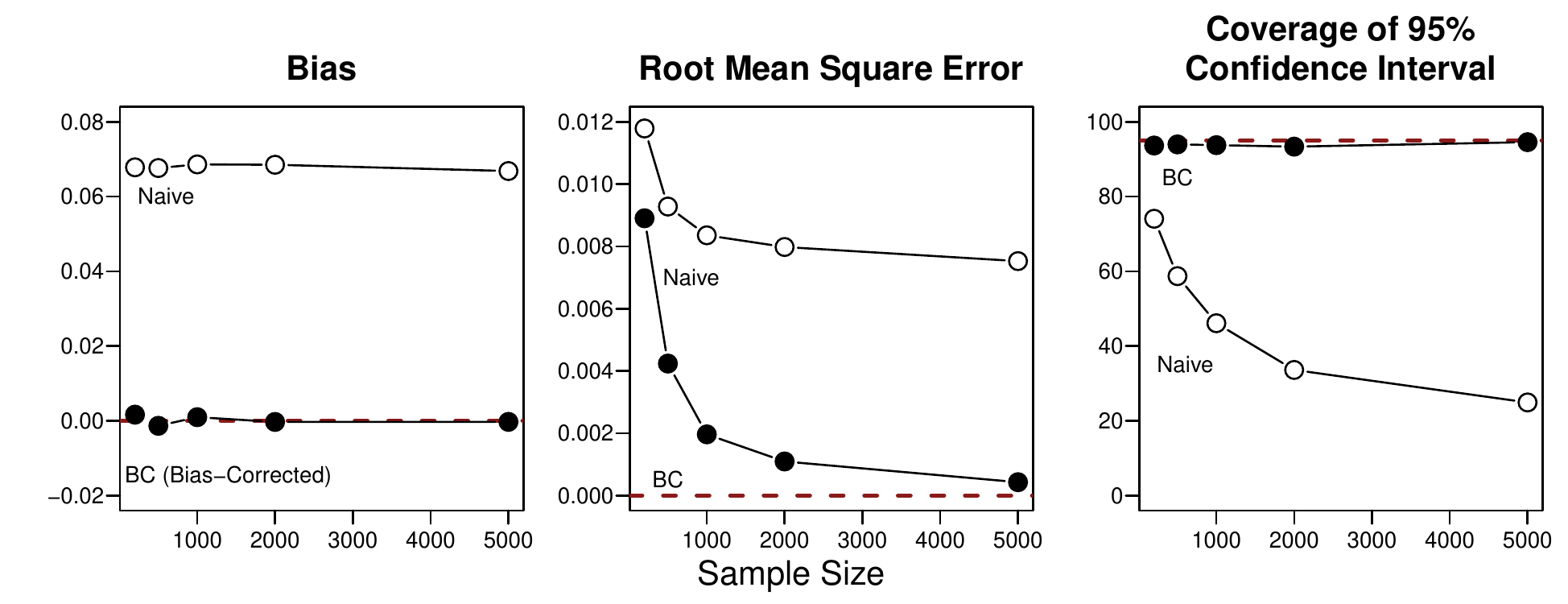}
\caption{ \textbf{Finite Sample Performance of the Na\"ive and Bias-Corrected Estimators.} 
 \textit{Note}: This figure displays the bias, root-mean-square-error, and the coverage of the 95\% confidence interval of the na\"ive and bias-corrected estimators. The bias-corrected estimator is unbiased and consistent and has an ideal level of coverage.}\label{fig:simulate2}
\end{figure}

To examine the finite sample performance of the bias-corrected estimator, we replicate the simulations that appeared in Figure \ref{fig:simulate0} 8000 times. In each simulation, we draw $\pi$ from the continuous uniform distribution (0.1, 0.45), $p$ and $p'$ from the continuous uniform distribution (0.088, 0.333) (reflecting  the smallest and largest values in existing studies), and $\gamma$ from the continuous uniform distribution (0.5, 1). Finally, we repeat the simulations for different sample sizes of 200, 500, 1000, 2000, and 5000 and evaluate the results. Figure \ref{fig:simulate2} demonstrates that the bias-corrected estimator has a significantly lower bias, smaller root-mean-square error (RMSE), and higher coverage than the na\"ive estimator.  


\begin{figure}[tbh!]
    \centering
   \includegraphics[width=12cm]{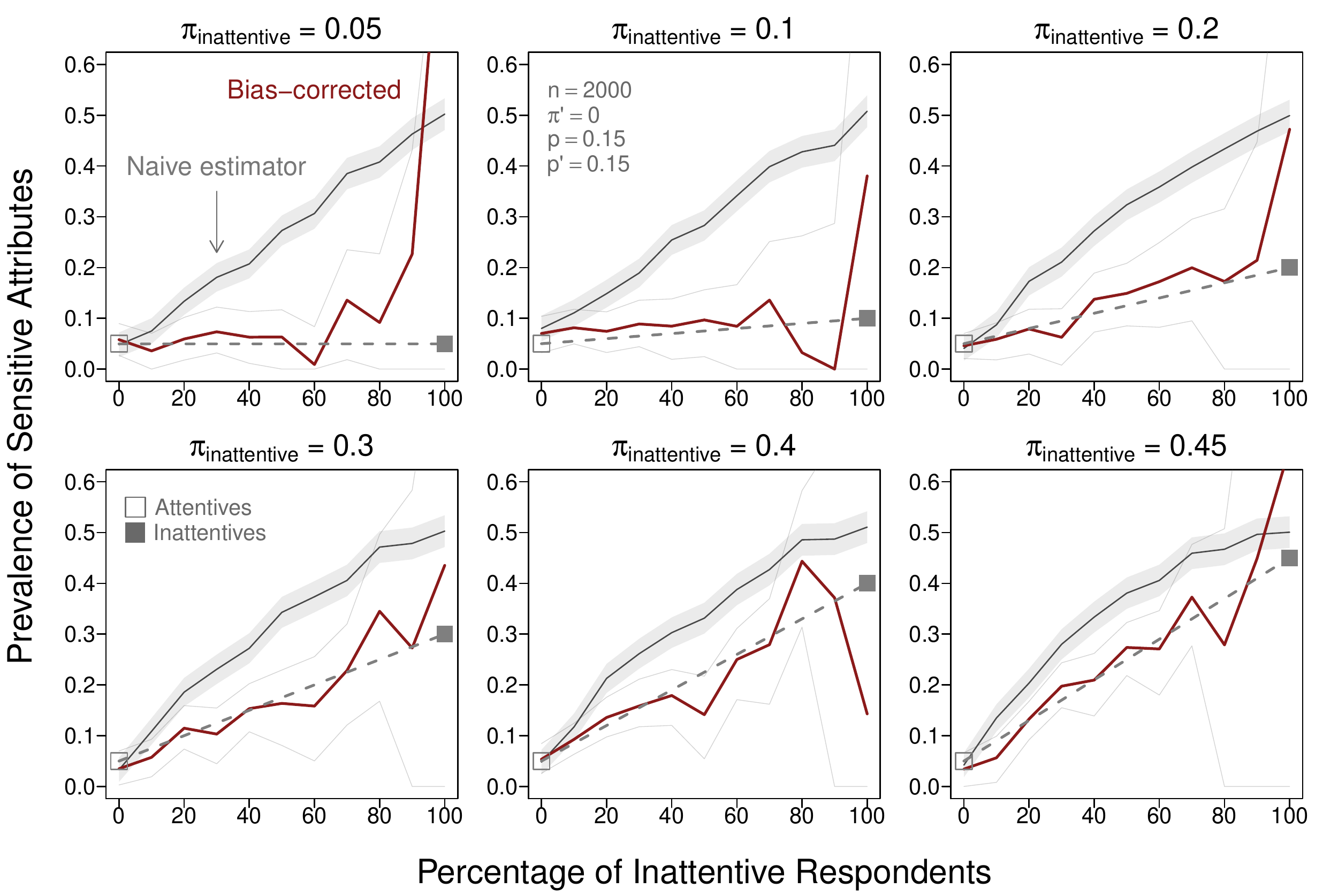}
\caption{\textbf{When Respondents with Sensitive Attributes Tend to Be More Inattentive.} \textit{Note}: This graph illustrates the na\"ive and bias-corrected estimators with 95\% confidence intervals when the prevalence of sensitive attributes among inattentive respondents ($\pi_{\text{inattentive}}$) is higher than that among attentive respondents ( $\pi_{\text{attentive}}$) with simulated data (see the top-middle panel for parameter values). The bias-corrected estimator captures the ground truth (dashed line) even when respondents with sensitive attributes tend to be more inattentive.}\label{fig:simulate3}
\end{figure}

The recent survey literature suggests that researchers be cautious when inattentive respondents have different outcome profiles (e.g., political interest) than attentive respondents \citep{berinsky2014separating,alvarez2019paying}. Paying attention to this issue is especially important when dealing with sensitive questions since respondents who have sensitive attributes may be more likely to give random answers than other respondents. To investigate whether our correction is robust to such an association, we further replicate our simulations in Figure \ref{fig:simulate0} by varying the prevalence of sensitive attributes among inattentive respondents (0.05, 0.1, 0.2,0.3,0.4,0.45) while holding the prevalence among attentive respondents constant (0.05). Figure \ref{fig:simulate3} show that our bias-corrected estimator properly captures the true prevalence rate of sensitive attributes regardless of the degree of association between inattentiveness and possession of sensitive attributes. In contrast, the na\"ive estimator never captures the ground truth when more than 10\% of respondents are inattentive. More simulation results are reported in Section \ref{sec:appSimulation}.

\section{Extensions of the Bias-Corrected Estimator}\label{sec:extension}
\subsection{Sensitivity Analysis}
\begin{figure}[tbh!]
    \centering
    \includegraphics[width=14cm]{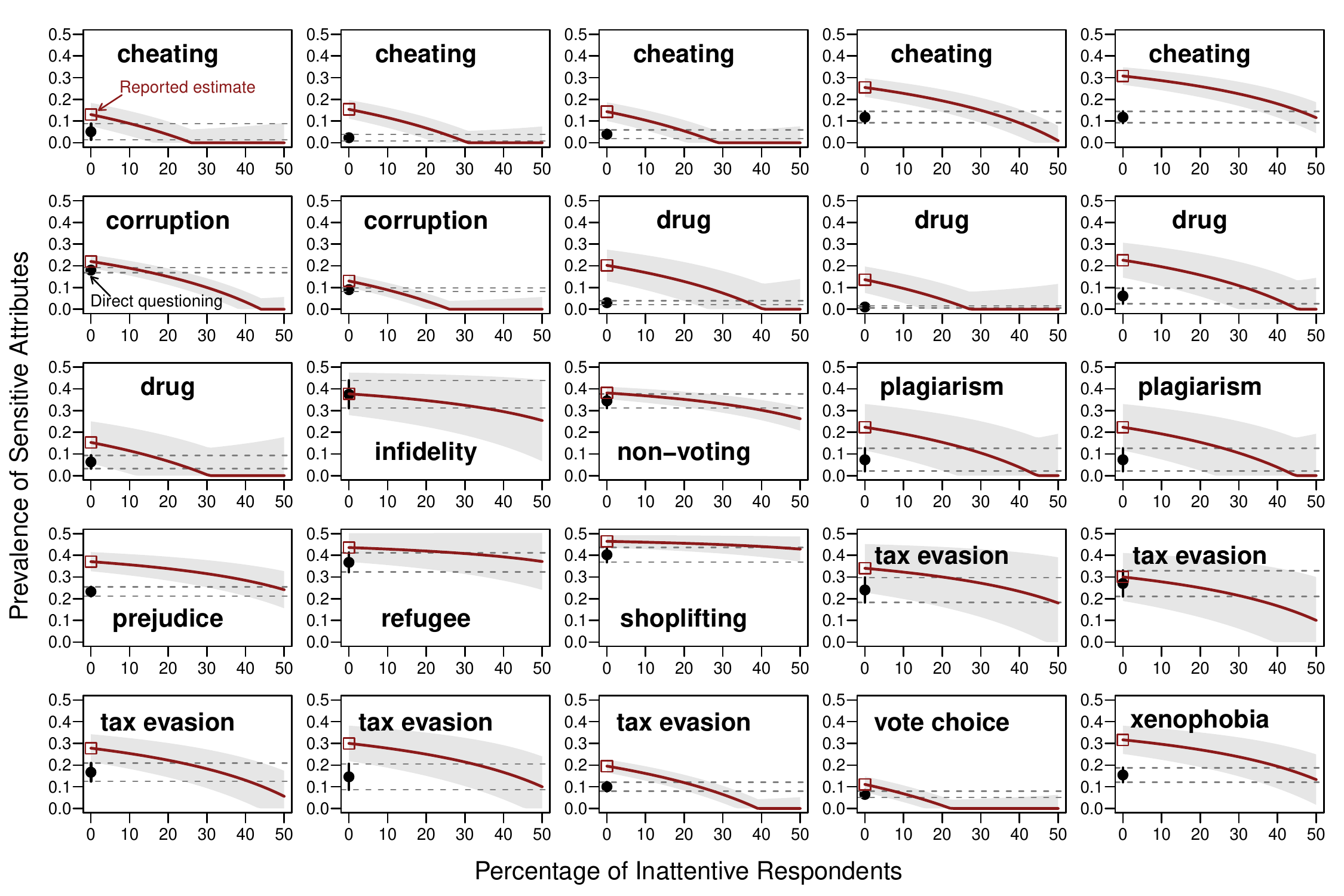}
    \flushleft
    \caption{{\bf Sensitivity Analysis of Previous Crosswise Estimates.} \textit{Note}: This figure plots bias-corrected estimates of the crosswise model over varying percentages of inattentive respondents with the estimate based on direct questioning reported in each study.}
    \label{fig:sensitivity}
\end{figure}

While our bias-corrected estimator requires the anchor question, it may not always be available (e.g, in existing crosswise surveys). For such surveys, we propose a sensitivity analysis that shows researchers the sensitivity of their na\"ive estimates to inattentive respondents and what assumptions they must make to preserve their original conclusions. Specifically, it offers sensitivity bounds for original crosswise estimates by applying the bias correction under varying levels of inattentive respondents. To illustrate this procedure, we attempted to apply our sensitivity analysis to all published crosswise studies of sensitive behaviors from 2008 to the present (49 estimates reported in 21 original studies). Figure \ref{fig:sensitivity} visualizes the sensitivity bounds for selected studies (see Section \ref{app:sensitivity} for full results).\footnote{The selected studies have more than 200 respondents, point estimates between 0.05 and 0.5, and direct questioning that appears to underestimate the quantity of interest.} For each study, we plot the bias-corrected estimates against varying percentages of inattentive respondents under Assumption 1. Our sensitivity analysis implies that many studies would not find any statistically significant difference between direct questioning and the crosswise model \citep[echoing][135]{hoglinger2017uncovering} unless they assume that less than 20\% of the respondents were inattentive.\footnote{Our software also allows researchers to set values of $\kappa$ other than 0.5 depending on the nature of their surveys.\label{foonote}}

\subsection{Weighting}\label{sec:weighting}
While the literature on the crosswise model usually assumes that survey respondents are drawn from a finite target population via simple random sampling with replacement, a growing share of surveys are administered with unrepresentative samples such as online opt-in samples \citep{franco2017developing,mercer2018weighting}. Online opt-in samples are known to be often unrepresentative of the entire population that researchers want to study \citep{malhotra2007effect,bowyer2017mode} and analysts using such samples may wish to use weighting to extend their inferences into the population of real interest. In this light, we propose a simple way to include sample weights in the bias-corrected estimator. Section \ref{app:weighting} presents our theoretical and simulation results. The key idea is that we can apply a Horvitz-Thompson-type estimator \citep{horvitz1952generalization} to the observed proportions in the crosswise and anchor questions (for a similar result without bias correction, see \citet[380]{chaudhuri2012unbiased} and \citet[270]{quatember2019discussion}). 

\subsection{Multivariate Regressions for the Crosswise Model}
Political scientists often wish to not only estimate the prevalence of sensitive attributes (e.g., corruption in a legislature), but also analyze what kinds of individuals (e.g., politicians) are more likely to have the sensitive attribute and probe whether having the sensitive attribute is associated with another outcome (e.g., reelection). The main challenge in drawing such inferences under the crosswise model is that analysts do not observe any individual-level information about sensitive attributes. Nevertheless, we demonstrate that it is possible to study these individual-level associations in a regression framework with only aggregate-level information about sensitive attributes, along with individual-level covariates, while correcting for inattention bias.

Similar regression models for the traditional crosswise model have been proposed in several studies \citep{jann2011asking,vakilian2014estimation,korndorfer2014measuring,gingerich2016protect}. Our contribution is to further extend such a framework by (1) enabling analysts to use the latent sensitive attribute \textit{both} as an outcome and a predictor while (2) applying our bias correction. Our software can easily implement these regressions while also offering simple ways to perform post-estimation simulation (e.g., generating predicted probabilities with 95\% confidence intervals).

\subsubsection{Using the Latent Sensitive Attribute as an Outcome}
We first introduce multivariate regressions in which the latent (unobserved) variable for having a sensitive attribute is used as an outcome variable. Let $Z_{i}\in\{0,1\}$ be a binary variable denoting if respondent $i$ has a sensitive attribute and $T_{i}\in\{0,1\}$ be a binary variable denoting if the same respondent is attentive. Both of these quantities are unobserved \textit{latent} variables. We define the regression model (conditional expectation) of interest as:
\begin{subequations}
\begin{align}
   \E[Z_{i}|\mathbf{X_{i}=x}] = \mathbb{P}(Z_{i}=1|\mathbf{X_{i}=x}) = \pi_{\bm{\beta}}(\mathbf{x}) \label{eq:Z},
\end{align}

\noindent where $\mathbf{X_{i}}$ is a random vector of respondent $i$'s characteristics, $\mathbf{x}$ is a vector of specific values of such covariates, and $\bm{\beta}$ is a vector of unknown parameters that associate these characteristics with the probability of having the sensitive attribute. Our goal is to make inferences about these unknown parameters and to use estimated coefficients to produce predictions for interpretation.

To apply our bias correction, we also introduce the following conditional expectation for being attentive:
\begin{equation}
    \E[T_{i}|\mathbf{X_{i}=x}] = \mathbb{P}(T_{i}=1|\mathbf{X_{i}=x}) = \gamma_{\bm{\theta}}(\mathbf{x})\label{eq:T},
\end{equation}
\end{subequations}

\noindent where $\bm{\theta}$ is a vector of unknown parameters that associate the same respondent's characteristics with the probability of being attentive. We then assume that $\pi_{\bm{\beta}}(\mathbf{x}) = \text{logit}^{-1}(\bm{\beta} \mathbf{X_{i}})$ and $\gamma_{\bm{\theta}}(\mathbf{x}) = \text{logit}^{-1}(\bm{\theta} \mathbf{X_{i}})$, acknowledging that other functional forms are also possible.

Next, we substitute these quantities into Equation (\ref{eq:lambda}) by assuming $\pi'=0$ (zero-prevalence for \textsf{Statement C}):
\begin{subequations}
\begin{align}
\lambda_{\bm{\beta,\theta}}(\mathbf{X_{i}}) & =       \underbrace{\Big(\pi_{\bm{\beta}}(\mathbf{X_{i}}) p + (1-\pi_{\bm{\beta}}(\mathbf{X_{i}}))(1-p)\Big) \gamma_{\bm{\theta}}(\mathbf{X_{i}}) + \frac{1}{2}\Big(1-\gamma_{\bm{\theta}}(\mathbf{X_{i}})\Big)}_{\text{Conditional probability of choosing ``both or neither is true'' in the crosswise question}}\label{eq:lambdaBeta}\\
\lambda'_{\bm{\theta}}(\mathbf{X_{i}}) & =      \underbrace{ \Big( \frac{1}{2}-p' \Big)\gamma_{\bm{\theta}}(\mathbf{X_{i}}) + \frac{1}{2}}_{\text{Conditional probability of choosing ``both or neither is true'' in the anchor question}}\label{eq:lambdaTheta}
\end{align}
\end{subequations}

Finally, let $Y_{i}\in\{0,1\}$ and $A_{i}\in\{0,1\}$ be observed variables denoting if respondent $i$ chooses ``both or neither is true'' in the crosswise and anchor questions, respectively. Assuming $Y_{i}\indep A_{i}|\mathbf{X_{i}}$ with Assumptions 1-3, we model that $Y_{i}$ and $A_{i}$ follow independent Bernoulli distributions with success probabilities $\lambda_{\bm{\beta,\theta}}(\mathbf{X_{i}})$ and $\lambda'_{\bm{\theta}}(\mathbf{X_{i}})$ and construct the following likelihood function:
\begin{align*}
    \mathcal{L}(\bm{\beta}, \bm{\theta}|\{\mathbf{X_{i}}, Y_{i}, A_{i}\}_{i=1}^{n}, p, p') & =  \prod_{i=1}^{n}  \Big\{ \lambda_{\bm{\beta,\theta}}(\mathbf{X_{i}}) \Big\}^{Y_{i}} \Big\{ 1 - \lambda_{\bm{\beta,\theta}}(\mathbf{X_{i}}) \Big\}^{1-Y_{i}} \Big\{ \lambda'_{\bm{\theta}}(\mathbf{X_{i}}) \Big\}^{A_{i}}\Big\{ 1 -  \lambda'_{\bm{\theta}}(\mathbf{X_{i}}) \Big\}^{1 - A_{i}}\\ 
     & =  \prod_{i=1}^{n}  \Big\{ \Big(\pi_{\bm{\beta}}(\mathbf{X_{i}}) p + (1-\pi_{\bm{\beta}}(\mathbf{X_{i}}))(1-p)\Big) \gamma_{\bm{\theta}}(\mathbf{X_{i}}) + \frac{1}{2}\Big(1-\gamma_{\bm{\theta}}(\mathbf{X_{i}})\Big)  \Big\}^{Y_{i}}\\
    & \quad \ \ \ \times \Big\{ 1 - \Big[\Big(\pi_{\bm{\beta}}(\mathbf{X_{i}}) p + (1-\pi_{\bm{\beta}}(\mathbf{X_{i}}))(1-p)\Big) \gamma_{\bm{\theta}}(\mathbf{X_{i}}) + \frac{1}{2}\Big(1-\gamma_{\bm{\theta}}(\mathbf{X_{i}})\Big) \Big] \Big\}^{1-Y_{i}}\\
    & \quad \ \ \ \times \Big\{  \Big( \frac{1}{2}-p' \Big)\gamma_{\bm{\theta}}(\mathbf{X_{i}}) + \frac{1}{2} \Big\}^{A_{i}}\\
    & \quad \ \ \ \times \Big\{ 1 - \Big[ \Big( \frac{1}{2}-p' \Big)\gamma_{\bm{\theta}}(\mathbf{X_{i}}) + \frac{1}{2} \Big] \Big\}^{1 - A_{i}}\\ \color{black}    
    & =  \prod_{i=1}^{n}  \Big\{ \Big((2p-1)\pi_{\bm{\beta}}(\mathbf{X_{i}}) + \Big(\frac{1}{2}-p \Big)\Big) \gamma_{\bm{\theta}}(\mathbf{X_{i}}) + \frac{1}{2} \Big\}^{Y_{i}}\\
    & \quad \ \ \ \times \Big\{ 1 - \Big[ \Big((2p-1)\pi_{\bm{\beta}}(\mathbf{X_{i}}) + \Big(\frac{1}{2}-p \Big)\Big) \gamma_{\bm{\theta}}(\mathbf{X_{i}}) + \frac{1}{2} \Big] \Big\}^{1-Y_{i}}\\
    & \quad \ \ \ \times \Big\{  \Big( \frac{1}{2}-p' \Big)\gamma_{\bm{\theta}}(\mathbf{X_{i}}) + \frac{1}{2} \Big\}^{A_{i}}\\
    & \quad \ \ \ \times \Big\{ 1 - \Big[ \Big( \frac{1}{2}-p' \Big)\gamma_{\bm{\theta}}(\mathbf{X_{i}}) + \frac{1}{2} \Big] \Big\}^{1 - A_{i}}
    \numberthis \label{eq:Likelihood}
\end{align*}

Our simulations show that estimating the above model can recover both primary ($\bm{\beta}$) and auxiliary parameters ($\bm{\theta}$) (Section \ref{app:regress}).\footnote{For estimation, we take a natural log of this likelihood function and maximize it by an iterative maximization method. For inference, we use the negative inverse of the Hessian matrix of the log-likelihood evaluated at the maximum likelihood estimates to compute standard errors of the estimated coefficients. The same applies to the next regression model.} 

\subsubsection{Using the Latent Sensitive Attribute as a Predictor}
Next, we propose regressions models in which the latent sensitive attribute is used as a predictor or explanatory variable. Let $V_{i}$ be a continuous or discrete outcome variable for respondent $i$ (other types of outcome variables are also possible). We define the regression model (conditional expectation) of interest as:
\begin{subequations}
\vspace{-0.5cm}
\begin{align}
   g_{\bm{\Theta}}(V_{i}|\mathbf{X_{i}}, Z_{i})\label{V}, 
\end{align}



\noindent where $\bm{\Theta}$ is a vector of parameters that associate a set of predictors plus the indicator for having a sensitive trait ($\mathbf{X_{i}}$, $Z_{i}$) and the response variable ($V_{i}$). For example, for a normally distributed outcome variable, we can consider $g_{\bm{\Theta}}(V_{i}|\mathbf{X_{i}}, Z_{i}) = \mathcal{N}(\alpha+\bm{\gamma}^{\top}\mathbf{X_{i}} + \delta Z_{i}, \sigma^2)$ with $\bm{\Theta}=(\alpha, \bm{\gamma}, \delta, \sigma^2)$. For a binary response variable, we can consider $g_{\bm{\Theta}}(V_{i}|\mathbf{X_{i}}, Z_{i}) = \text{Bernoulli}(\phi)$, where $\frac{\phi}{1-\phi} = \alpha+\bm{\gamma}^{\top}\mathbf{X_{i}} + \delta Z_{i}$ and $\bm{\Theta}=(\alpha, \bm{\gamma}, \delta)$. Our goal is to make inferences about the association between the latent sensitive attribute ($Z_{i}$) and the response variable ($V_{i}$) after controlling for other covariates ($\delta$ is our primary quantity of interest).



Using all the available information from data, the observed data likelihood function then becomes:
\begin{align*}
    \mathcal{L}(\bm{\beta}, \bm{\theta}, \bm{\Theta}|\{V_{i},\mathbf{X_{i}}, Y_{i}, A_{i}\}_{i=1}^{n}, p, p') & =  \prod_{i=1}^{n}  g_{\bm{\Theta}}(V_{i}|\mathbf{X_{i}}, Z_{i}, T_{i})\mathbb{P}(Y_{i}=1,Z_{i},T_{i}|\mathbf{X_{i}})\mathbb{P}(A_{i}=1,Z_{i},T_{i}|\mathbf{X_{i}})\\
    & =  \prod_{i=1}^{n} \Bigg\{
    g_{\bm{\Theta}}(V_{i}|\mathbf{X_{i}}, 1, 1)p^{Y_{i}}(1-p)^{1-Y_{i}}\pi_{\bm{\beta}}(\mathbf{X_{i}})(1-p')^{A_{i}}p'^{1-A_{i}}\gamma_{\bm{\theta}}(\mathbf{X_{i}})\\
    & \ \qquad + g_{\bm{\Theta}}(V_{i}|\mathbf{X_{i}}, 0, 1)(1-p)^{Y_{i}}p^{1-Y_{i}}(1-\pi_{\bm{\beta}}(\mathbf{X_{i}}))(1-p')^{A_{i}}p'^{1-A_{i}}\gamma_{\bm{\theta}}(\mathbf{X_{i}}) \\
    & \ \qquad + g_{\bm{\Theta}}(V_{i}|\mathbf{X_{i}}, 1, 0)\frac{1}{2}\pi_{\bm{\beta}}(\mathbf{X_{i}})\frac{1}{2}(1-\gamma_{\bm{\theta}}(\mathbf{X_{i}}))\\
    & \ \qquad  + g_{\bm{\Theta}}(V_{i}|\mathbf{X_{i}}, 0, 0)\frac{1}{2}(1-\pi_{\bm{\beta}}(\mathbf{X_{i}}))\frac{1}{2}(1-\gamma_{\bm{\theta}}(\mathbf{X_{i}})) \Bigg\}, 
    \numberthis \label{eq:Likelihood3}
\end{align*}
\end{subequations}

\noindent where each part inside the bracket is $g_{\bm{\Theta}}(V_{i}|\mathbf{X_{i}}, z, t)\mathbb{P}(Y_{i}=1|Z_{i}=z, T_{i}=t)\mathbb{P}(Z_{i}=z|\mathbf{X_{i}})\mathbb{P}(A_{i}=1|Z_{i}=z,T_{i}=t)\mathbb{P}(T_{i}=1|\mathbf{X_{i}})$, where $z=\{0,1\}$ and $t=\{0,1\}$.

We assume that Assumptions 1-3 hold as well as $V_{i}\indep Y_{i}|\mathbf{X_{i}}$, $V_{i}\indep A_{i}|\mathbf{X_{i}}$, and $Y_{i}\indep A_{i}|\mathbf{X_{i}}$. The key insight is that, under these assumptions, we can rewrite the entire likelihood of the observed crosswise data as a product of three conditional probabilities. We can then marginalize the product over the two latent variables $Z_{i}$ and $T_{i}$ by summing up the conditional probabilities that we could in principle obtain for all possible combinations of the latent variables.\footnote{For example, the third component inside the bracket represents $g_{\bm{\Theta}}(V_{i}|\mathbf{X_{i}}, 1, 0)\mathbb{P}(Y_{i}=1|Z_{i}=1, T_{i}=0)\mathbb{P}(Z_{i}=1|\mathbf{X_{i}})\mathbb{P}(A_{i}=1|Z_{i}=1,T_{i}=0)\mathbb{P}(T_{i}=1|\mathbf{X_{i}})$. Here, $\mathbb{P}(Y_{i}=1|Z_{i}=1, T_{i}=0)$ is the conditional probability that respondents choose the crosswise item when they actually have a sensitive trail \textit{and} do not provide attentive responses. Because they do not follow the instruction, Assumption 1 states that this probability is $\frac{1}{2}$ (regardless of $Z_{i}$). Next, $\mathbb{P}(Z_{i}=1|\mathbf{X_{i}})$ is the conditional probability that respondents have a sensitive trait, and we defined this quantity as $\pi_{\bm{\beta}}(\mathbf{X_{i}})$. Now, $\mathbb{P}(A_{i}=1|Z_{i}=1,T_{i}=0)$ is the conditional probability that respondents choose the crosswise item in the anchor question when they actually have a sensitive trait \textit{and} do not provide attentive responses. Because they do not follow the instruction, Assumption 1 states that this probability is $\frac{1}{2}$ (regardless of $Z_{i}$). Finally, $\mathbb{P}(T_{i}=1|\mathbf{X_{i}})$ is the conditional probability that respondents \textit{do not} provide attentive responses, and we defined this quantity as $1 - \gamma_{\bm{\theta}}(\mathbf{X_{i}})$. Hence, the joint probability for this component is $ g_{\bm{\Theta}}(V_{i}|\mathbf{X_{i}}, 1, 0)\frac{1}{2}\pi_{\bm{\beta}}(\mathbf{X_{i}})\frac{1}{2}(1-\gamma_{\bm{\theta}}(\mathbf{X_{i}}))$.} Our simulation studies confirm that this regression model can recover both the main ($\bm{\Theta}$) and auxiliary ($\bm{\beta}, \bm{\theta}$) parameters (Section \ref{app:regress_pred}). We plan to demonstrate the effectiveness of the proposed regression frameworks with empirical examples in future research.

\subsection{Sample Size Determination and Parameter Selection}

When using the crosswise model with our procedure, researchers may wish to choose the sample size and specify other design parameters (i.e., $\pi'$, $p$, and $p'$) so that they can obtain (1) high statistical power for hypothesis testing and/or (2) narrow confidence intervals for precise estimation. To fulfill these needs, we develop power analysis and data simulation tools appropriate for our bias-corrected estimator.

First, our power analysis uses a one-sided hypothesis test based on the Wald test \citep{ulrich2012asking}. We consider the null hypothesis $H_{0}:\pi \leq \pi_{0},$ where $\pi_{0}$ (prevalence rate under the null) may be zero or a particular value obtained from direct questioning. Assuming that the crosswise estimate is larger than the direct questioning estimate --- and larger than zero (i.e., more-is-better assumption), we then consider the alternative hypothesis $H_{1}: \pi > \pi_{0}$ when the true value of $\pi$ is $\pi_{1}$. Based on the normal approximation, the power function becomes:
\begin{align}
\underbrace{\mathbb{P}(\{ \text{Reject $H_{0} | H_{1}$ is true} \})}_{\text{Power}} = \beta = 1 - \underbrace{\Phi\Big(\frac{\pi_{0} - \pi_{1} - c \tilde{\sigma}_{0}}{\tilde{\sigma}_{1}}\Big)}_{\text{$\mathbb{P}$(Type II Error)}},
\end{align}

\noindent where $\Phi$ is the cumulative distribution function of the standard normal distribution, $c = \Phi(1-\alpha)$ is the critical value given size $\alpha =  \mathbb{P}(\text{Type I Error})$, and $\tilde{\sigma}_{0}$ and $\tilde{\sigma}_{1}$ are \textit{simulated} standard errors of the bias-corrected estimates under $H_{0}$ and $H_{1},$ respectively.\footnote{We simulate the standard errors in repeated Monte Carlo experiments at $n=\{1,500,1000,1500,2000,2500\}$. Our software also allows one to draw more fine-grained power curves.}

\begin{figure}[tbh!]
    \centering
    \includegraphics[width=14cm]{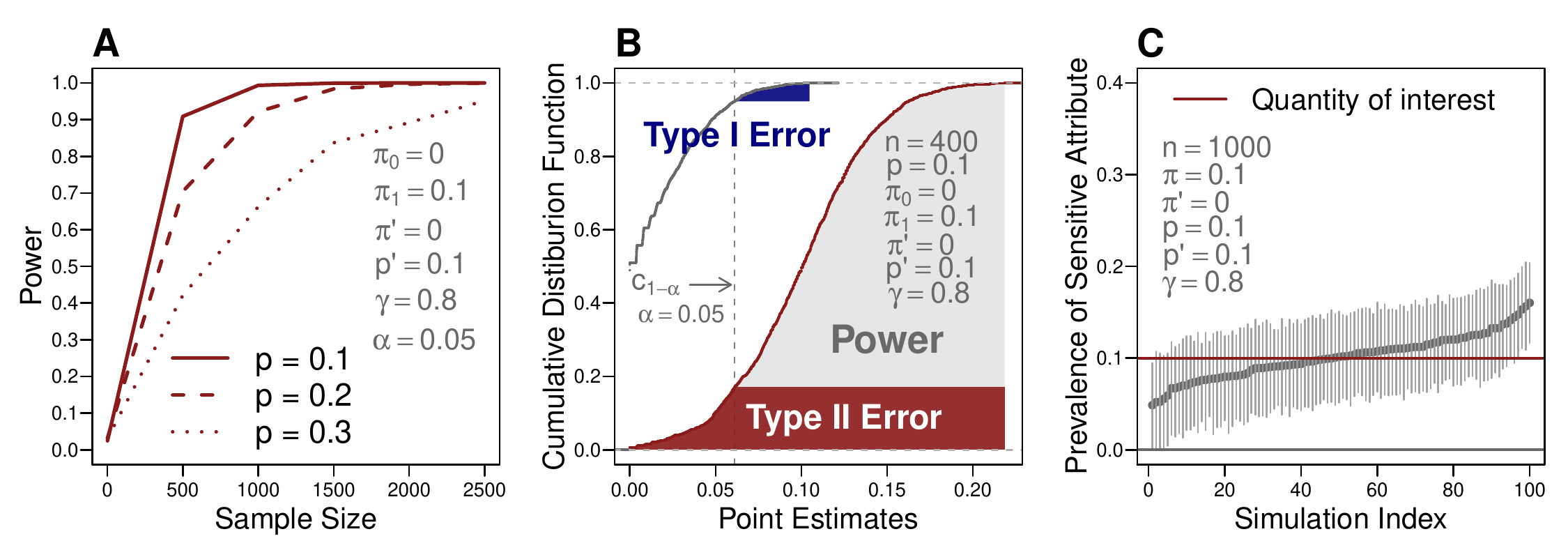}
    \flushleft
    \caption{{\bf Tools for Power Analysis and Parameter Selection.} \textit{Note}: Panel A shows power curves for three different values of $p$. Panel B offers visualizes type I error, type II error, and power as areas under the empirical cumulative distribution of point estimates under $H_{0}$ (left) and $H_{1}$ (right). Panel C display 100 point estimates and 95\% confidence intervals based on 100 simulated data sets.}
    \label{fig:power}
\end{figure}

Panel A of Figure \ref{fig:power} plots power curves against the sample sizes. For this illustration, we assume that $\pi_{0}=0$, $\pi_{1}=0.1$, $\pi'=0$,  $p'=0.1$, $\gamma=0.8$, and $\alpha=0.05$. Substantively, this means that we hope to distinguish the estimated prevalence of 0.1 from zero with a 95\% (100\% $\times$ $1-\alpha$) level of confidence when we expect that only 80\% of respondents are attentive. Panel A displays multiple power curves for different values of $p$. It suggests that if researchers want to reject $H_{0}$ with power $\beta=0.8$ they need to have about 400 (when $p=0.1$), 600 (when $p=0.2$), and 1400 (when $p=0.3$) respondents.

After choosing the sample size $n$, researchers may wish to verify if the selected $n$ and design parameters achieve the desired level of power and re-adjust their design parameters if necessary. To further assist applied researchers, Panel B visualizes type I error, type II error, and statistical power as areas under the (empirical) cumulative distribution function of the sampling distribution of the bias-corrected estimates based on $H_{0}$ (left) and $H_{1}$ (right). We assume the same parameter values as above and $n=400$ and $p=0.1$. Panel B shows a small type I error and high power (close to 0.8) or equivalently small type II error as we expected.
 
Next, our data simulation tool allows researchers to see what they would obtain from using the crosswise model with some fixed sample size and parameter values. The key idea is to give analysts a sense of how large their 95\% confidence interval would be and let them change the sample size and design parameters to achieve their desired precision (i.e., desired length of the confidence interval). To illustrate, Panel C shows the point estimates and 95\% confidence intervals based on 100 simulated data sets with $n=1000$ and $\pi=0.1$, $\pi'=0$, $p=p'=0.1$, $\gamma=0.8$. It suggests that the resulting interval estimate would be approximately the point estimate $\pm 0.05$. If a narrower confidence interval is needed, researchers should increase the sample size and/or choose lower values for $p$ and $p'$.

\section{Concluding Remarks and Practical Suggestions}\label{sec:guide}

The crosswise model is a simple but powerful survey-based tool for investigating the prevalence of sensitive attitudes and behavior. To overcome two limitations of the design, we proposed a simple design-based solution using an anchor question. We also provided several extensions of our proposed bias-corrected estimator, which allow researchers to analyze sensitive topics with the crosswise model in ways that were not available before. Future research could further extend our methodology by applying it to non-binary sensitive questions, by allowing for multiple sensitive statements and/or anchor questions, by handling missing data more efficiently, by using the latent sensitive attribute for covariate balance, and by integrating our method with more efficient sampling schemes \citep[e.g.,][]{reiber2020improving}.

With these developments, we hope to facilitate the wider adoption of the (bias-corrected) crosswise design in political science, since, as discussed above, it has several potential advantages over traditional randomized response questions, list experiments, and endorsement experiments. Future research, however, should explore how to compare and combine results from the crosswise model and these other techniques. 

Furthermore, our work also speaks to recent scholarship on inattentive respondents in self-administered surveys. While it is increasingly common for political scientists to use self-administered surveys, the most common way of dealing with inattentive respondents is to try to directly identify them through ``Screener''-type questions. However, a growing number of respondents to surveys are experienced survey-takers who may recognize such traps and avoid them \citep{alvarez2019paying}. Our method is one example of a different approach to handling inattentive respondents that, with well-constructed anchor questions, is not likely to be recognized by respondents, works without measuring individual-level inattentiveness, and does not require excluding these respondents from the sample. Future research should explore whether a similar approach to inattention would work in other question formats.

Finally, we conclude with some practical advice for researchers when designing a bias-corrected crosswise model. First, to satisfy the random pick assumption (Assumption 1), researchers should randomize the order of the two crosswise answer-choices both in the crosswise and anchor questions. This can be done at the question level or (if there are many crosswise questions presented sequentially) at the respondent level (to avoid respondent confusion). Second, attention consistency (Assumption 2) is satisfied when the crosswise and anchor questions have the same \textit{proportion} of attentive respondents. If respondents, on average, perceive these two types of questions to be somehow different, attention consistency may be violated. Thus, we recommend that researchers design the anchor questions to fit in topically with the other crosswise questions in the survey, have a similar level of sensitivity, and look quite similar to them (e.g., the anchor question should use the same format for the non-sensitive item as the other crosswise questions (e.g., a mother's birthday) and the sensitive item should be about the same length in both the anchor and other crosswise questions). We discuss more of the subtleties of writing good anchor questions (and provide examples) in Section \ref{sec:appAnchor}. In addition, researchers may also compare the duration of time spent on the crosswise and anchor questions as a way of diagnosing any differences in how respondents have treated them. Finally, randomizing the position of the anchor question in the survey relative to the crosswise questions may help guarantee that there is no carryover effect from one type of question to another (Assumption 3).


\singlespacing
\bibliographystyle{apsr}
\bibliography{crosswise.bib}

\clearpage
\appendix

\singlespacing
\setcounter{page}{1} 
\part{Online Appendix} 
{{\LARGE For ``A Bias-Corrected Estimator for the Crosswise Model with Inattentive Respondents''}}\\

\vspace{0.1cm}
\noindent {\Large Yuki Atsusaka and Randolph T. Stevenson}\\

\vspace{0.1cm}

\parttoc 
\renewcommand\thefigure{\thesection.\arabic{figure}}
\renewcommand\thetable{\thesection.\arabic{table}}
\renewcommand\theequation{\thesection.\arabic{equation}}

\clearpage

\section{Additional Discussion on the Bias-Corrected Estimator}\label{sec:appDerivation}
\setcounter{figure}{0} 
\setcounter{table}{0}  
\setcounter{equation}{0} 
\setcounter{footnote}{0} 

\subsection{Derivation of the Bias}\label{sec:bias}

Here, we derive the bias in the na\"ive crosswise estimator based on the argument in Section 3. According to the conventional definition of bias in estimators, we define bias as the (signed) difference between the expected value of the na\"ive crosswise estimator and the true quantity of interest:
\begin{subequations}
\begin{align}
   B_{CM} & \equiv \E[\widehat\pi_{CM}] - \pi  \\
  & =  \E \Bigg[\frac{\widehat\lambda + p - 1}{2p - 1}\Bigg] - \frac{\lambda - (1-p)\gamma - \kappa(1-\gamma)   }{(2p - 1)\gamma} \\
  & = \frac{\gamma(\lambda+p-1) - ( \lambda - \gamma + p\gamma - \kappa + \kappa\gamma)}{(2p-1)\gamma}\\
  & = \frac{\lambda\gamma+p\gamma-\gamma - \lambda + \gamma - p\gamma + \kappa - \kappa\gamma}{(2p-1)\gamma}\\
  & = \frac{\lambda\gamma - \kappa\gamma -\lambda + \kappa}{(2p-1)\gamma}\\
  & = \frac{\lambda-\kappa}{(2p-1)} - \frac{\lambda - \kappa}{(2p-1)\gamma}\\
  & = \frac{1}{2}\Bigg( \frac{\lambda - \kappa}{p-\frac{1}{2}} \Bigg) - \frac{1}{2\gamma}\Bigg( \frac{\lambda - \kappa}{p-\frac{1}{2}}  \Bigg)\\
  & = \Bigg( \frac{1}{2} - \frac{1}{2\gamma} \Bigg)\Bigg( \frac{\lambda - \kappa}{p-\frac{1}{2}} \Bigg)
\end{align}
\end{subequations}

The second term of the second line was obtained by transforming Equation (1c) as:
\begin{subequations}
\begin{align}
 \lambda & = \Big\{ \pi p + (1-\pi)(1-p) \Big\}\gamma + \kappa(1-\gamma)\\
\Rightarrow \Big\{ \pi p + (1-\pi)(1-p) \Big\}\gamma & = \lambda -  \kappa(1-\gamma)\\
\Rightarrow \Big\{\pi p + 1 - p - \pi + \pi p\Big\}\gamma & = \lambda -  \kappa(1-\gamma)\\
\Rightarrow \Big\{ 2\pi p - \pi + 1 - p  \Big\}\gamma & = \lambda -  \kappa(1-\gamma)\\
\Rightarrow \Big\{ \pi(2p - 1) + 1 - p  \Big\}\gamma & = \lambda -  \kappa(1-\gamma)\\
\Rightarrow \pi(2p - 1)\gamma + (1 - p)\gamma & = \lambda -  \kappa(1-\gamma)\\
\Rightarrow \pi(2p - 1)\gamma  & = \lambda - (1 - p)\gamma - \kappa(1-\gamma)\\
\Rightarrow \pi  & = \frac{\lambda - (1 - p)\gamma - \kappa(1-\gamma)}{(2p - 1)\gamma }\\
\end{align}
\end{subequations}

\subsection{Behavior of the Bias}\label{sec:appBias}
By definition, the bias vanishes when the proportion of attentive respondents is 1 ($\gamma=1)$. To see this, simply observe the following limit:
\begin{subequations}
\begin{align}
 &   \lim_{\gamma \rightarrow 1}\Bigg( \frac{1}{2} - \frac{1}{2\gamma} \Bigg)\Bigg( \frac{\lambda - \kappa}{p-\frac{1}{2}} \Bigg)\\
 & \ \ \ =   \Bigg( \frac{1}{2} - \frac{1}{2} \Bigg)\Bigg( \frac{\lambda - \kappa}{p-\frac{1}{2}} \Bigg) \\
 & \ \ \ =   0
\end{align}
\end{subequations}

In contrast, as the proportion of attentive respondents approaches 0 (from the side of 1), the bias term explodes and approaches positive infinity. To see this, observe that the multiplier \big($\frac{1}{2} - \frac{1}{2\lambda}$\big) is always negative and the multiplicand \bigg($\frac{\lambda - \kappa}{p-\frac{1}{2}}$\bigg) is also negative under some regularity conditions. These conditions are that $\pi < 0.5$, and $p < \frac{1}{2}$ (and thus $\lambda > \kappa$). These regularity conditions hold in most surveys that use the crosswise model. Since the absolute value of the multiplier grows as $\lambda$ approaches 0, the bias term increases as the proportion of attentive responses decreases.

However, the limit itself does not exist as:
\begin{subequations}
\begin{align}
 &   \lim_{\lambda \rightarrow 0}\Bigg( \frac{1}{2} - \frac{1}{2\gamma} \Bigg)\Bigg( \frac{\lambda - \kappa}{p-\frac{1}{2}} \Bigg)\\
 & =   \textsf{Undefined}
\end{align}
\end{subequations}

\subsection{Simulation of the Bias}\label{sec:appBiasSim}

\begin{figure}[tbh!]
\centering 
\includegraphics[width=11cm]{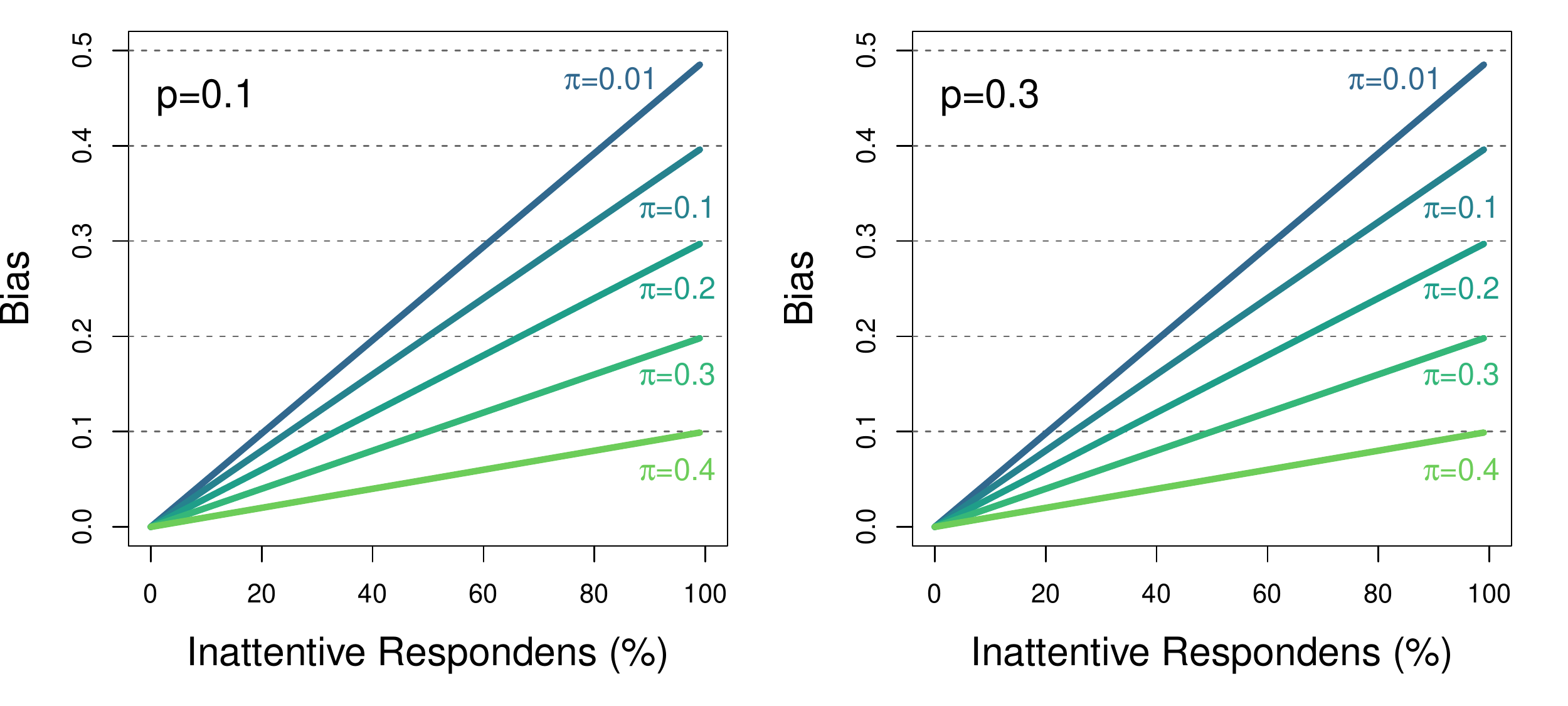}
\caption{\textbf{Bias in the Na\"ive Crosswise Estimator.} \textit{Note}: Both panels display the (theoretical) bias in the conventional crosswise estimator with varying levels of inattentive respondents.}\label{fig:simulate_bias}
\end{figure}

Figure \ref{fig:simulate_bias} visualizes the properties of the bias in the conventional crosswise estimator (assuming $\kappa=0.5$). It shows that the size of the bias increases as (1) the percentage of inattentive respondents increases and (2) the quantity of interest approaches 0, but that it does not change regardless of the value of $p$.

\subsection{Derivation of the Variance}\label{sec:variance}
Here, we derive the population and sample variance of the bias-corrected crosswise estimator discussed in Section 3.
\begin{subequations}
\begin{align}
  \V(\widehat \pi_{BC}) & = \V \Bigg[ \widehat\pi_{CM} - \Bigg( \frac{1}{2} - \frac{1}{2\widehat{\gamma}'} \Bigg)\Bigg( \frac{\widehat{\lambda} - \frac{1}{2}}{p-\frac{1}{2}} \Bigg) \Bigg]\\
& = \V \Bigg[ \widehat\pi_{CM} - \Bigg( \frac{1}{2} - \frac{1}{2}\Bigg[\frac{\frac{1}{2}-p'}{\widehat{\lambda}' - \frac{1}{2}}\Bigg] \Bigg)\Bigg( \frac{\widehat{\lambda} - \frac{1}{2}}{p-\frac{1}{2}} \Bigg) \Bigg]\\  
  & = \V \Bigg[ \frac{\widehat{\lambda}+p-1}{2p-1} - \Bigg( \frac{1}{2} - \frac{1}{2}\Bigg[\frac{\frac{1}{2}-p'}{\widehat{\lambda}' - \frac{1}{2}}\Bigg] \Bigg)\Bigg( \frac{\widehat{\lambda} - \frac{1}{2}}{p-\frac{1}{2}} \Bigg) \Bigg]\\
& = \V \Bigg[ \frac{\widehat{\lambda}+p-1}{2p-1} - \frac{1}{2}\Bigg( \frac{\widehat{\lambda} - \frac{1}{2}}{p-\frac{1}{2}} \Bigg) + \frac{1}{2}\Bigg( \frac{\widehat{\lambda} - \frac{1}{2}}{p-\frac{1}{2}} \Bigg) \Bigg(\frac{\frac{1}{2}-p'}{\widehat{\lambda}' - \frac{1}{2}} \Bigg) \Bigg]\\ & = \V \Bigg[ \frac{\widehat{\lambda}+p-1}{2p-1} - \frac{\widehat{\lambda}-\frac{1}{2}}{2p-1} + \Bigg(\frac{\widehat{\lambda}-\frac{1}{2}}{2p-1}  \Bigg) \Bigg(\frac{\frac{1}{2}-p'}{\widehat{\lambda}' - \frac{1}{2}} \Bigg) \Bigg]\\
& = \V \Bigg[ \frac{p-\frac{1}{2}}{2p-1}  +\Bigg(\frac{\widehat{\lambda}-\frac{1}{2}}{2p-1}  \Bigg) \Bigg(\frac{\frac{1}{2}-p'}{\widehat{\lambda}' - \frac{1}{2}} \Bigg) \Bigg]\\
& = \V \Bigg[ \frac{p-\frac{1}{2}}{2p-1}\Bigg]  + \V \Bigg[\Bigg(\frac{\widehat{\lambda}-\frac{1}{2}}{2p-1}  \Bigg) \Bigg(\frac{\frac{1}{2}-p'}{\widehat{\lambda}' - \frac{1}{2}} \Bigg) \Bigg]\\
& = 0  + \V \Bigg[\Bigg(\frac{\widehat{\lambda}-\frac{1}{2}}{2p-1}  \Bigg) \Bigg(\frac{\frac{1}{2}-p'}{\widehat{\lambda}' - \frac{1}{2}} \Bigg) \Bigg]\\
& = \V \Bigg[\Bigg(\frac{\widehat{\lambda}-\frac{1}{2}}{2p-1}  \Bigg) \Bigg(\frac{\frac{1}{2}-p'}{\widehat{\lambda}' - \frac{1}{2}} \Bigg) \Bigg]\\
& = \V \Bigg[\Bigg(\frac{\widehat{\lambda}-\frac{1}{2}}{\widehat{\lambda}' - \frac{1}{2}}  \Bigg) \Bigg(\frac{\frac{1}{2}-p'}{2p-1} \Bigg) \Bigg]\\
& = \V \Bigg[\Bigg(\frac{\widehat{\lambda}}{\widehat{\lambda}'}  \Bigg) \Bigg(\frac{\frac{1}{2}-p'}{2p-1} \Bigg) \Bigg]\\
\end{align}

Similarly, 
\begin{align}
  \widehat{\V}(\widehat \pi_{BC})  = \widehat{\V} \Bigg[\Bigg(\frac{\widehat{\lambda}}{\widehat{\lambda}'}  \Bigg) \Bigg(\frac{\frac{1}{2}-p'}{2p-1} \Bigg) \Bigg]\\
\end{align}
\end{subequations}

\subsection{Unbiasedness of the Na\"ive Estimator}\label{sec:naive}
To see that the na\"ive estimator is unbiased when $\gamma=1$,  let $Y_{i}$ be a binary random variable denoting whether respondent $i$ chooses the crosswise item (i.e., TRUE-TRUE or FALSE-FALSE) and its realization $y_{i} \in \{0,1\}$. Let the number of respondents choosing the crosswise item be $k = \sum_{i=1}^{N}y_{i}$, where $k < n$. Then, the likelihood function for $\lambda$ given any observed $k$ is $L(\lambda|n, k) = {n\choose k} \lambda^{k}(1-\lambda)^{n-k}$. Applying the first-order condition yields a maximum likelihood estimate (MLE) of $\lambda$, $ \widehat\lambda = \frac{k}{n}$, where $\E[\widehat\lambda] = \lambda.$ The unbiasedness follows from the fact that $\E[\widehat\lambda] =  \E\big[\frac{k}{n}\big] = \frac{1}{n}\E[k] = \frac{1}{n}n\lambda = \lambda$. Following the parameterization invariance property of MLEs, $\E[\widehat \pi_{CM}$] = $\pi$. This result, however, does not hold when $\gamma\neq1$.

\subsection{Unbiasedness of $\widehat{\lambda'}$}\label{sec:lambdaprime}

To see that $\widehat{\lambda'}$ is an unbiased estimator of $\lambda'$, let us define that $\widehat\lambda'$ is a binomial random variable (like $\widehat\lambda)$ with parameters $n, \lambda'$ and $\widehat\lambda' = k'/n$, where $k'$ is the number of people who choose the crosswise item in the anchor question. This is because $k' \sim \text{Binom}(n,\lambda')$ and $\widehat\lambda' = k'/n$ suggests $\widehat\lambda' \sim \text{Binom}(n,k').$ The probability mass function that $\widehat\lambda'$ taking $n'/n$ is given by $\text{Pr}(\widehat\lambda'= \frac{n'}{n}) = {n \choose n'}(k')^{n'}(1-k')^{n-n'}$.

\subsection{Unbiasedness of the Proposed Estimator}\label{sec:proofunbiased}

Here, we offer proof that the proposed estimator is an unbiased estimator of the quantity of interest under two assumptions.
\begin{subequations}
\begin{align}
    \E[\widehat{\pi}_{BC}] & =     \E \Bigg[\widehat{\pi}_{CM} -  \Bigg( \frac{1}{2} - \frac{1}{2}\Bigg[\frac{\frac{1}{2}-p'}{\widehat{\lambda}' - \frac{1}{2}}\Bigg] \Bigg)\Bigg( \frac{\widehat\lambda - \frac{1}{2}}{p-\frac{1}{2}} \Bigg) \Bigg] \\
     & =     \E [\widehat{\pi}_{CM}] -  \E\Bigg[\Bigg( \frac{1}{2} - \frac{1}{2}\Bigg[\frac{\frac{1}{2}-p'}{\widehat{\lambda}' - \frac{1}{2}}\Bigg] \Bigg)\Bigg( \frac{\widehat\lambda - \frac{1}{2}}{p-\frac{1}{2}} \Bigg) \Bigg] \\
& =    \E [\widehat{\pi}_{CM}]  -  \E\Bigg[\Bigg( \frac{1}{2} - \frac{1}{2\widehat{\gamma}'} \Bigg)\Bigg( \frac{\widehat\lambda - \frac{1}{2}}{p-\frac{1}{2}} \Bigg) \Bigg] \\  
& =    \E [\widehat{\pi}_{CM}]  -  \Bigg( \frac{1}{2} - \frac{1}{2\E[\widehat{\gamma}']} \Bigg)\Bigg( \frac{\E[\widehat\lambda] - \frac{1}{2}}{p-\frac{1}{2}} \Bigg)  \\ 
& =    \E [\widehat{\pi}_{CM}] -  \Bigg( \frac{1}{2} - \frac{1}{2\gamma'} \Bigg)\Bigg( \frac{\E[\widehat\lambda] - \frac{1}{2}}{p-\frac{1}{2}} \Bigg)  \\ 
& =    \E [\widehat{\pi}_{CM}] -  \Bigg( \frac{1}{2} - \frac{1}{2\gamma} \Bigg)\Bigg( \frac{\E[\widehat\lambda] - \frac{1}{2}}{p-\frac{1}{2}} \Bigg) \quad \text{(by Assumption 2)}  \\ 
& =   \E [\widehat{\pi}_{CM}] -  \Bigg( \frac{1}{2} - \frac{1}{2\gamma} \Bigg)\Bigg( \frac{\lambda - \frac{1}{2}}{p-\frac{1}{2}} \Bigg) \quad \text{(by Assumption 3)} \\ 
& =   \E [\widehat{\pi}_{CM}] -  B_{CM}  \\ 
& =  \pi  \quad \text{(by the definition of the bias)}
\end{align}
\end{subequations}

\clearpage
\section{Discussions and Evaluations of Alternative Estimators}\label{sec:appSimulation}
\setcounter{figure}{0} 
\setcounter{table}{0}  
\setcounter{equation}{0} 
\setcounter{footnote}{0} 

In this appendix section, we describe two other methods for adjusting crosswise estimates for bias resulting from respondent inattention. In the first subsection we describe these methods a explain why we should only expect them to be useful under ideal conditions unlikely to be met in most settings in which the crosswise model would be used. Next, we present a set of simulation based comparison of the bias, root mean squared error, and coverage of the 95\% confidence intervals for these estimators against our own.  

\subsection{Description of Alternative Estimators Estimator}

 The first suggestion for correcting bias due to inattention in the crosswise estimator was by \citetSM{enzmann2017}, who made the suggestion in a book chapter on crosswise methods, but provided no discussion or justification for his proposal. His proposal begins with the formula below and then suggests estimating the unknown proportion of inattentive respondents by asking, in an unprotected direct question, whether the respondent answered the crosswise question randomly. It's unclear from his brief treatment whether this should be done separately for each crosswise question in a survey or only once for the survey as a whole.   
 
\begin{subequations}
\begin{align}
 \frac{\widehat{\pi}_{CM} - 0.5r}{1-r}\label{eq:SchnappEnzman}    
\end{align}
\end{subequations}

\noindent where $\widehat{\pi}_{CM}$ is the naive crosswise estimator and $r$ is the population proportion of inattentive respondents. 

In a later paper, \citetSM[311]{schnapp2019sensitive} reproduces Enzmann’s formula and then leans into the idea of estimating the percentage of random responders in this equation via direct questioning. Further, he adds an equation for the variance of the estimator. Below, we call this the Enzmann-Schnapp estimator, or just ``ES.''  

Besides describing the Enzmann method, \citetSM{schnapp2019sensitive}  also presents his own proposal for a bias-corrected estimator, which he calls the CMR-I. That estimator also asks respondents if they answered the crosswise question(s) randomly and then uses that answer to adjust individual answers to the crosswise questions accordingly (e.g., changing the answers for respondents that report answering randomly from either the recorded 0 or 1 to 0.5 and then using all the adjusted and unadjusted responses together to calculate $\widehat{\lambda}$ (the proportion of respondents choosing ``both or neither is true'') in the usual crosswise formula). He then applies the usual variance estimator for crosswise models (not the one he proposed for Enzmann's estimator) to this estimator. 

The main difference in our estimator and the ES and CMR-I estimators is that, while we use anchor questions to estimate the share of inattentive respondents, these estimators both do so by directly asking respondents whether they answered the crosswise question(s) randomly. And, of course, the CMR-I estimator also differs more fundamentally in its overall construction. Likewise, Schnapp's variance estimate for this estimator is derived under the assumption that the percentage of random responders is known and so is not appropriate for the full estimator he actually describes (and uses), which includes the use of direct questioning about inattention.

Importantly, in both the ES and CMR-I estimators, the direct question about random responding is asked without the protection of the crosswise format and so it is likely that at least some respondents will hesitate to admit to that socially-undesirable behavior. Thus, in our simulations below, we present results for both The ES and CMR-I estimators with varying levels of veracity for the direct question about random guessing.

\subsection{Comparative Evaluation of the Estimators}

To perform our comparative evaluation, we replicate the simulations presented in Figure 1. To reiterate, in each simulation, we draw $\pi$ from continuous uniform distribution (0.1, 0.45), $p$ and $p'$ from continuous uniform distribution (0.088, 0.333) (reflecting  the smallest and largest values in existing studies), and $\gamma$ from continuous uniform distribution (0.5, 1). Finally, we repeat the set of experiments for different sample sizes of 200, 500, 1000, 2000, and 5000 and evaluate the results. To include the ES and CMR-I estimators in the simulations, we simulate direct questioning that asks whether respondents give random answers in the crosswise question with three different percentages of ``lairs'' in the direct question: 0\%, 25\%, and 50\%. We call these conditions ``No lairs'', ``25\% lairs'', and ``50\% lairs'', respectively. When simulating respondent answers to the direct question about whether they answered randomly, we also make the following assumption:\\

\noindent \textbf{Assumption B.1 (One-side Lying).} \textit{Only inattentive respondents may lie about giving random answers in the crosswise questions.}\\

In other words, we assume that attentive respondents always report that they did not give random answers in the crosswise model.

\begin{figure}[tbh!]
\centering
 \includegraphics[width=17cm]{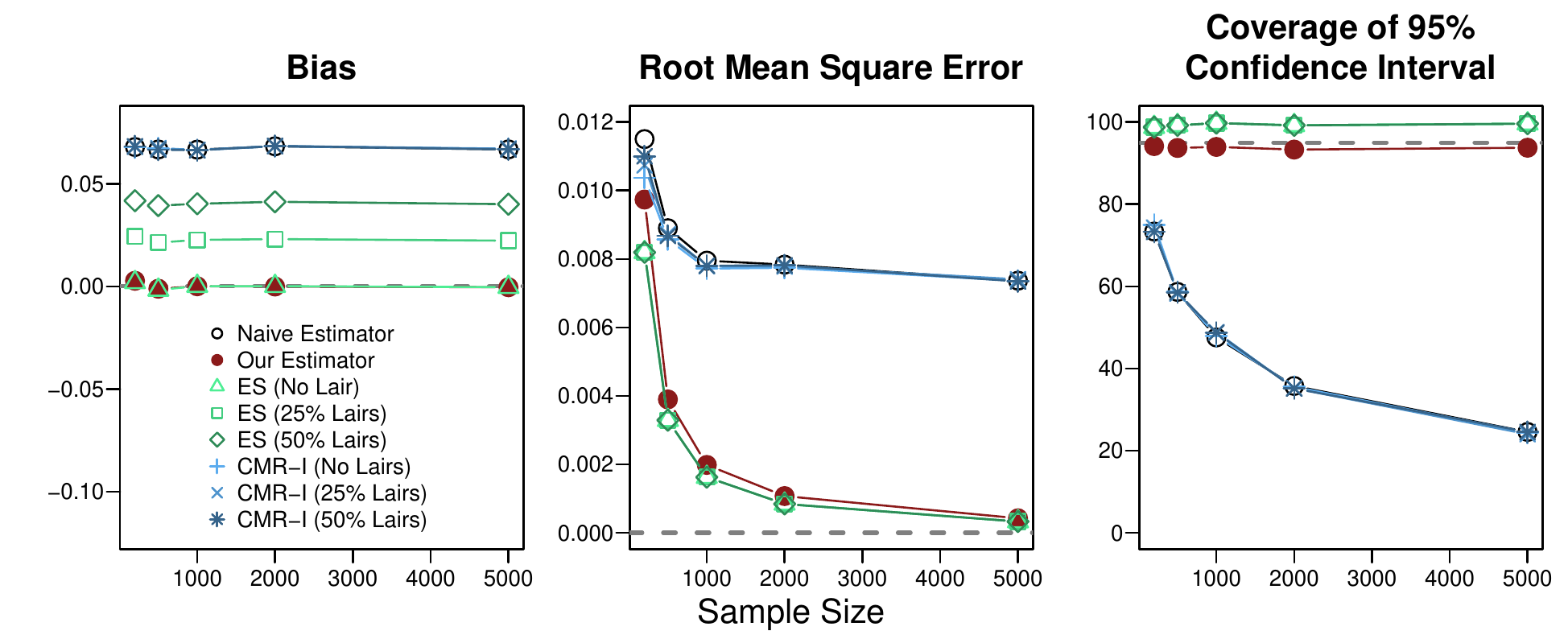}
\caption{ \textbf{Finite Sample Performance of the Na\"ive, Bias-Corrected, and Enzmann-Schnapp Estimators.} 
 \textit{Note}: This figure displays the bias, root-mean-square error, and the coverage of 95\% confidence interval of na\"ive estimator, our estimator, Enzmann-Schanpp correction (No Lairs),  Enzmann-Schanpp correction (25\% Lairs),  Enzmann-Schanpp correction (50\% Lairs), CMR-I (No Lairs),  CMR-I (25\% Lairs), and CMR-I (50\% Lairs). }\label{fig:simulate_biasEnzmann}
\end{figure}

Figure \ref{fig:simulate_biasEnzmann} report the results of our comparative assessments. It suggests that the CMR-I estimator does not work at all and the ES estimator only works under the no lair condition, which is hard to satisfy. Additionally, the ES estimator appears to have a wider coverage of the ground truth than necessary (i.e., their 95\% confidence intervals appear to cover the ground truth almost 100\% of the time). In contrast, the 95\% confidence interval of our estimator covers the ground truth 95\% of the times as expected.

If one focuses only on the bias and MSE results, it is clear that the our estimator and the ES estimator produce identical results under the No Liars condition. Indeed, in Section B.3 below, we show that if (and only if) that assumption (and one other) holds, the ES estimator is algebraically equivalent to ours. Of course, this key assumption is unlikely to hold in any realistic setting.

\begin{figure}[tbh!]
\centering
 \includegraphics[width=17cm]{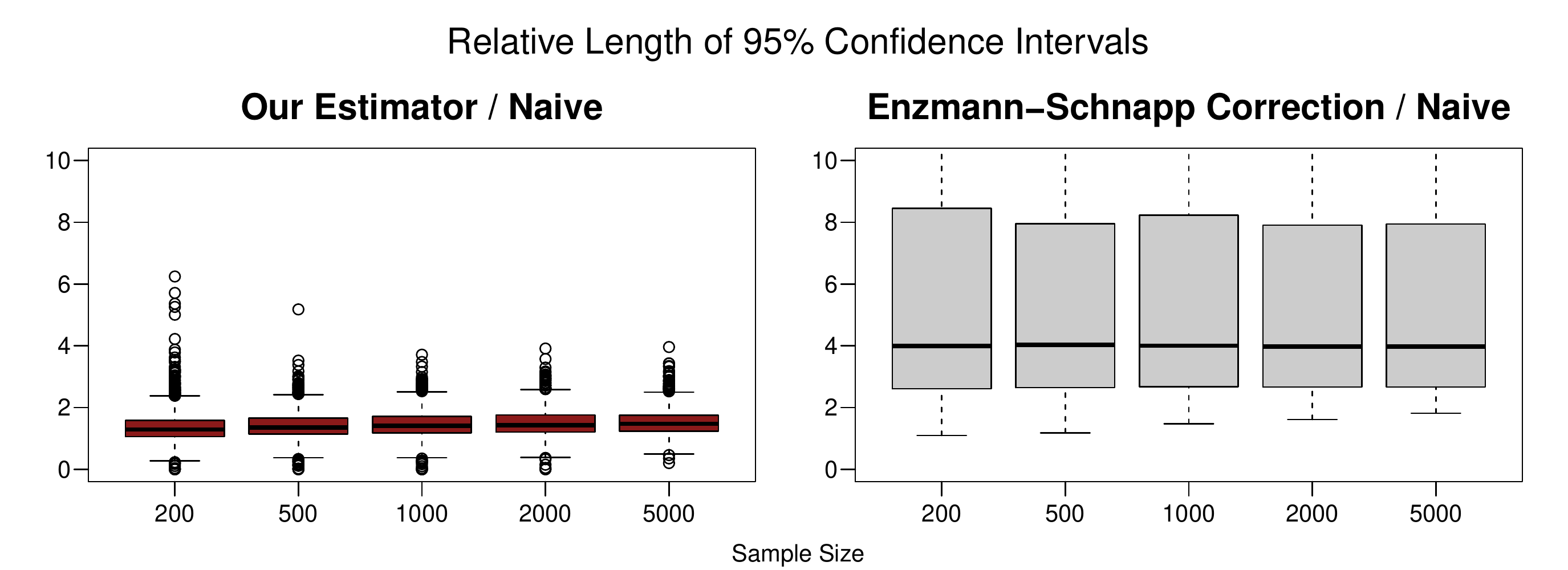}
\caption{ \textbf{Relative Lengths of 95\% Confidence Interval with respect to the Na\"ive estimator.} 
 \textit{Note}: This figure displays the length of the 95\% confidence intervals of each estimator relative to that of the conventional estimator on the $y$-axis. For example, for 200 observations, the confidence interval of the Enzmann-Schnapp correction is about 4 times larger than that of the na\"ive estimator.}\label{fig:RelativeEff}
\end{figure}

To further investigate the relative efficiency of our estimator vs. the ES in its most favorable condition (i.e., no lair), Figure \ref{fig:RelativeEff} shows the relative length of the 95\% confidence interval of our estimator (left panel) and the ES (right panel) to that of the conventional estimator. The figure illustrates that regardless of the sample size our estimator has a narrower confidence interval (about 1.3 times larger than the conventional crosswise confidence interval) than does the ES (about 4 times larger than the conventional crosswise confidence interval).

\newpage
\subsection{Proof that Equation (3d) and Equation (B.1.a) are Equivalent under Several Conditions}

In this section we show that our bias-corrected estimator is equivalent to the Enzmann-Schnapp correction if and only if we assume that $r$ is an \textit{estimated} proportion of inattentive respondents (i.e., $r = 1- \widehat{\gamma}$), Assumption 1 holds (i.e., $\kappa=0.5$), and the proportion of inattentive respondents is estimated in the same way using the anchor question. To prove the equivalence, we show that the difference between the two estimators is zero under these conditions.
\begin{subequations}
\begin{align}
\widehat{\pi}_{\text{Schnapp-Enznmann}} -\widehat{\pi}_{\text{BC}}  & = \frac{\widehat{\pi}_{CM} - 0.5r}{1-r} - \Bigg[ \widehat{\pi}_{CM} - \widehat B_{CM}  \Bigg]\\
& =\frac{\widehat{\pi}_{CM} - \frac{1}{2}(1-\widehat{\gamma})}{\widehat{\gamma}} - \widehat{\pi}_{CM} + \widehat B_{CM} \\
& = \frac{\widehat{\pi}_{CM}}{\widehat{\gamma}} - \frac{1}{2\widehat{\gamma}}(1-\widehat{\gamma}) - \widehat{\pi}_{CM}  + \widehat B_{CM}\\
& = \Bigg(\frac{1}{\widehat{\gamma}} - 1\Bigg)\widehat{\pi}_{CM}  - \Bigg(\frac{1}{\widehat{\gamma}} - 1\Bigg)\frac{1}{2}   + \widehat B_{CM}\\
& = \Bigg(\frac{1}{\widehat{\gamma}} - 1\Bigg)\Bigg(\widehat{\pi}_{CM}  - \frac{1}{2}\Bigg)   + \widehat B_{CM}\\
& = \Bigg(\frac{1}{\widehat{\gamma}} - 1\Bigg)\Bigg(\frac{\widehat\lambda + p - 1}{2p - 1}  - \frac{1}{2}\Bigg)   + \widehat B_{CM}\\
& = \Bigg(\frac{1}{\widehat{\gamma}} - 1\Bigg)\Bigg(\frac{\widehat\lambda + p - 1}{2(p - \frac{1}{2})}  - \frac{1}{2}\Bigg)   + \widehat B_{CM}\\
& = \Bigg(\frac{1}{\widehat{\gamma}} - 1\Bigg)\Bigg(\frac{\widehat\lambda + p - 1 - (p - \frac{1}{2})}{2(p - \frac{1}{2})}\Bigg)   + \widehat B_{CM}\\
& = \Bigg(\frac{1}{\widehat{\gamma}} - 1\Bigg)\Bigg(\frac{\widehat\lambda + p - 1 - p + \frac{1}{2}}{2(p - \frac{1}{2})}\Bigg)   + \widehat B_{CM}\\
& = \Bigg(\frac{1}{\widehat{\gamma}} - 1\Bigg)\Bigg(\frac{\widehat\lambda - \frac{1}{2}}{2(p - \frac{1}{2})}\Bigg)   + \widehat B_{CM}\\
& = \Bigg(\frac{1}{2\widehat{\gamma}} - \frac{1}{2}\Bigg)\Bigg(\frac{\widehat\lambda - \frac{1}{2}}{p - \frac{1}{2}}\Bigg)   + \widehat B_{CM}\\
& = - \Bigg(\frac{1}{2}-\frac{1}{2\widehat{\gamma}} \Bigg)\Bigg(\frac{\widehat\lambda - \frac{1}{2}}{p - \frac{1}{2}}\Bigg)   + \widehat B_{CM}\\
& = - \widehat B_{CM}   + \widehat B_{CM}\\
& = 0
\label{eq:Equivalence}    
\end{align}
\end{subequations}



\clearpage
\section{Additional Discussion and Simulations for Extensions}\label{sec:appExtension}
\setcounter{figure}{0} 
\setcounter{table}{0}  
\setcounter{equation}{0} 
\setcounter{footnote}{0} 

In this section, we present additional information for the proposed extensions of the bias-corrected estimator.




\subsection{Sensitivity Analysis}\label{app:sensitivity}

Here, we present the full results of our sensitivity analysis. We applied our sensitivity analysis to all existing studies (49 estimates reported by 21 unique articles) for which we were able to collect essential information. For consistency, if the original estimate ($\widehat{\pi}$) is between 0.5 and 1 (where we expect over-reporting), we flip the direction of the estimate so that it will be between 0 and 0.5 (where we expect under-reporting). We did not apply this transformation for Figure 5. The featured studies are from \citetSM{coutts2011plagiarism,jann2011asking,kundt2013re,korndorfer2014measuring,shamsipour2014estimating,kundt2014applying,hoffmann2015strong,gingerich2016protect,waubert2017indirect,hoglinger2018more,nasirian2018does,kuhn2018reducing,banayejeddi2019implementation,hoffmann2019prejudice,hopp2019estimating,klimas2019higher,meisters2020can,hoffmann2020validity,ozgul2020survey,mieth2021they,canan2021estimating}.

\begin{figure}[tbh!]
    \centering
    \includegraphics[width=15cm]{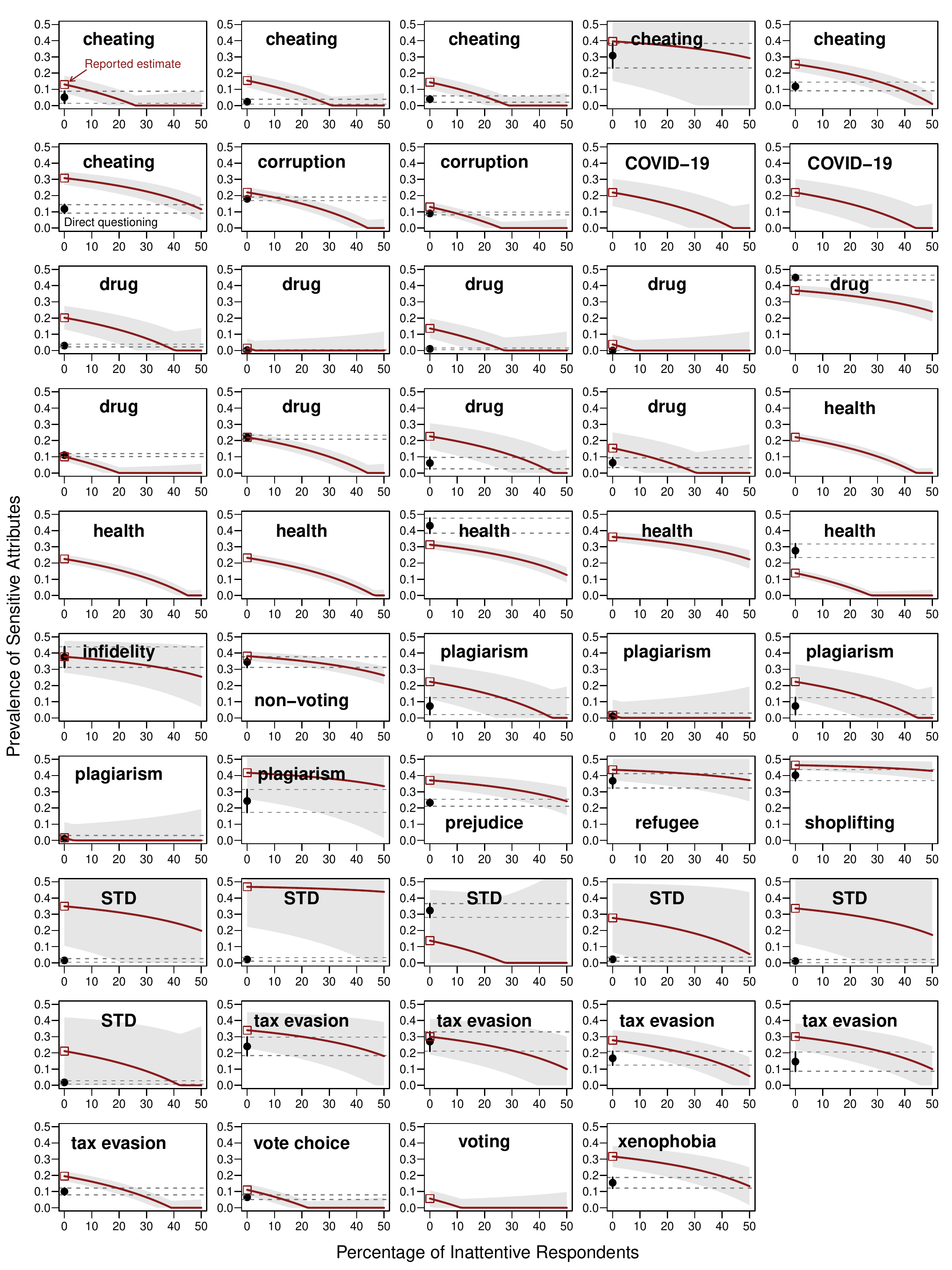}
    \caption{{\bf Sensitivity Analysis of Previous Crosswise Estimates.} \textit{Note}: For each estimate, the bias correction is applied with varying percentages of inattentive respondents.}
    \label{fig:sensitivity_full}
\end{figure}

\clearpage
\subsection{Weighting}\label{app:weighting}

Here, we present our strategy to incorporate sample weights into our bias-corrected estimator. Recall that the only sample statistics we observe in our framework are $\hat \lambda$ and $\hat \lambda'$, which are observed proportions of respondents choosing the crosswise item in the crosswise and anchor questions, respectively. The key idea here is that we can apply a Horvitz-Thompson-type estimator of the mean (and thus the inverse probability weighting more generally) to the crosswise proportions, where weights are the inverse of the probabilities that respondents in different strata will be in the sample. Namely, we can apply a weight $w_{i} = \frac{1}{\text{Pr}(S_{i}=1|\mathbf{X_{i}})},$ where $S_{i}=\{0,1\}$ is a binary variable denoting if respondent $i$ is in the sample and $\mathbf{X_{i}}$ is a vector of the respondent's background characteristics.\footnote{In this article, we only consider the base weight, but one can naturally include other weights such as non-response weights to construct the final survey weights.}

Let $Y_{i}\in\{0,1\}$ be a binary variable denoting if respondent $i$ chooses the crosswise item in the crosswise question and $A_{i}\in\{0,1\}$ be a binary variable denoting if the same respondent chooses the crosswise item in the anchor question. We then estimate the weighted crosswise proportion in the crosswise question and the weighted proportion of attentive respondents in the following way:
\begin{subequations}
\begin{align}
    \widehat\lambda_{w} & = \dfrac{\sum_{i=1}^{n}w_{i}Y_{i}}{ \sum_{i=1}^{n}w_{i}  } \quad \text{and} \quad
    \widehat\gamma_{w} = \dfrac{\dfrac{\sum_{i=1}^{n}w_{i}A_{i}}{  \sum_{i=1}^{n}w_{i}} - \frac{1}{2}}{\frac{1}{2}-p'},
\end{align}
\end{subequations}

The proof is straightforward. Assuming that $Y_{i}\indep S_{i}|X$ (choosing the crosswise item and being in the sample are statistically independent conditional upon a covaraite), weighting can recover the population crosswise proportion $\lambda$ from the sample crosswise response $Y_{i}S_{i}$:

\begin{align*}
  &   \E\Bigg[ \frac{Y_{i}S_{i}}{\text{Pr}(S_{i}=1|X)}\Bigg]\\
   = & \E \Bigg[ \E \Bigg[ \frac{Y_{i}S_{i}}{\text{Pr}(S_{i}=1|X)} \Bigg| X  \Bigg] \quad \text{(Iterative Expectation)}\\
  = & \E \Bigg[ \frac{ \E[Y_{i}|X]\E[S_{i}| X]}{\text{Pr}(S_{i}=1|X)}   \Bigg] \quad \text{(Conditional Independence)}\\
  = & \E \Bigg[ \frac{ \E[Y_{i}|X] \text{Pr}(S_{i}=1|X) }{\text{Pr}(S_{i}=1|X)}   \Bigg] \quad \text{(Definition of Expectation)}\\ 
  = & \E [ \E[Y_{i} | X] ]\\
  = & \E[Y_{i}] \quad \text{(Iterative Expectation)}\\
  = & \lambda
\end{align*}

Similarly, we can show that
\begin{align*}
  &   \E\Bigg[ \frac{A_{i}S_{i}}{\text{Pr}(S_{i}=1|X)}\Bigg]\\
   = & \E \Bigg[ \E \Bigg[ \frac{A_{i}S_{i}}{\text{Pr}(S_{i}=1|X)} \Bigg| X  \Bigg]\\
   = & \E[A_{i}]\\
   = & \lambda'
\end{align*}

In practice, researchers can calculate weights using their favorite weighting techniques such as raking (or iterative proportional fitting), matching, propensity score weighting, or sequential applications of these. Recent research shows that ``when it comes to accuracy, choosing the right variables for weighting is more important than choosing the right statistical method'' \citepSM[4]{mercer2018weighting}. Thus, we recommend that researchers think carefully about the association between the sensitive attribute of interest and basic demographic and other context-dependent factors when using weighting. To choose the ``right'' variables, our proposed regression models can also be useful exploratory aids. When generalizing the results on sensitive attributes to a larger population, however, it is strongly advised to elaborate on how weights are constructed and what potential bias may exist \citepSM{franco2017developing}.

Another possible approach to deal with highly selected samples is to employ multilevel regression and post-stratification (MRP) \citepSM{downes2018multilevel}. While we do not consider MRP with crosswise estimates in this article, future research should explore the optimal strategy to use MRP in sensitive questions.

To illustrate our weighting strategy, we simulate crosswise data with two covariates: $X_{1} \sim \text{Binomial}(0.5)$ and $X_{2} \sim \text{Poisson}(30)$ for, for example, 100,000 voters. Specifically, we simulate the true prevalence rates in the crosswise and anchor questions according to the following generative models:
\begin{align*}
 \pi & = \frac{\exp(\beta_{0} + \beta_{1}X_{1} + \beta_{2}X_{2})}{1 + \exp(\beta_{0} + \beta_{1}X_{1} + \beta_{2}X_{2})}\\
 \gamma & = \frac{\exp(\theta_{0} + \theta_{1}X_{1} + \theta_{2}X_{2})}{1 + \exp(\theta_{0} + \theta_{1}X_{1} + \theta_{2}X_{2})},
\end{align*}

\noindent where we set $\beta_{0}=-1.5$, $\beta_{1}=0.5$, $\beta_{2}=0.02$ and $\theta_{0}=2$, $\theta_{1}=-0.1$, $\theta_{2}=-0.01.$. For example, we can consider $X_{1}$ as a binary indicator for being female (as opposed to non-female) and female voters are more likely to have sensitive traits than non-female voters (i.e., $\beta_{1}=0.5)$.

Under this generative model, the population-level proportion of individuals with sensitive traits is \textbf{0.35} (and the population proportion is \textbf{0.40} for female and \textbf{0.29} for non-female voters). Now, from the population of 100,000 voters, we sample 1000, 2000, 3000, 4000, and 5000 individuals. In this process, we intentionally oversample female voters with probability 0.7. Consequently, we obtain sample weights 1.43 for female voters and 3.33 for non-female voters. We then generate bias-corrected crosswise estimates with and without incorporating the sample weights.

\begin{figure}[tbh!]
    \centering
    \includegraphics[width=12cm]{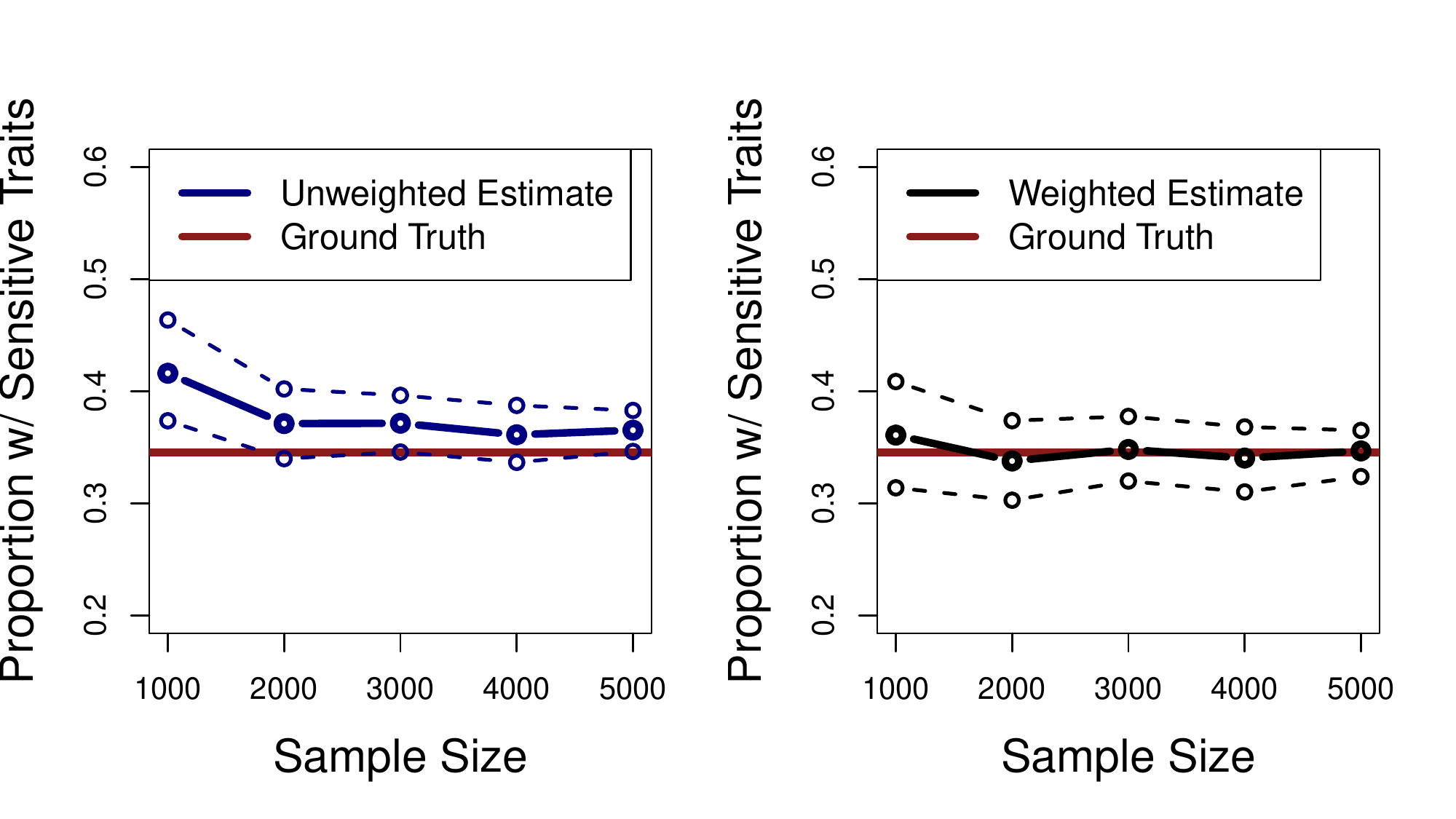}
\caption{ \textbf{Weighting in the Crosswise Estimator.} \textit{Note}: This figure shows bias-corrected crosswise estimates without weighting (left panel) and estimates with weighting (right panel) based on simulated unrepresentative samples.}\label{fig:weighting}
\end{figure}

Figure \ref{fig:weighting} compares bias-corrected crosswise estimates based on simulated unrepresentative samples with and without sample weights. It demonstrates that while the unweighted estimator always overestimates the ground truth, the weighted estimator captures the population-level quantity of interest.

\subsection{Multivariate Regression: Sensitive Attribute as an Outcome}\label{app:regress}

To validate this regression framework, we simulate crosswise data with two covariates: $X_{1} \sim \text{Binomial}(0.5)$ and $X_{2} \sim \text{Poisson}(30).$ Specifically, we simulate the true prevalence rates in the crosswise and anchor questions according to the following generative models:
\begin{align*}
 \pi & = \frac{\exp(\beta_{0} + \beta_{1}X_{1} + \beta_{2}X_{2})}{1 + \exp(\beta_{0} + \beta_{1}X_{1} + \beta_{2}X_{2})}\\
 \gamma & = \frac{\exp(\theta_{0} + \theta_{1}X_{1} + \theta_{2}X_{2})}{1 + \exp(\theta_{0} + \theta_{1}X_{1} + \theta_{2}X_{2})},
\end{align*}

\noindent where we set $\beta_{0}=-1.5$, $\beta_{1}=0.5$, $\beta_{2}=0.02$ and $\theta_{0}=2$, $\theta_{1}=-0.1$, $\theta_{2}=-0.01.$.

\begin{figure}[tbh!]
    \centering
    \includegraphics[width=14cm]{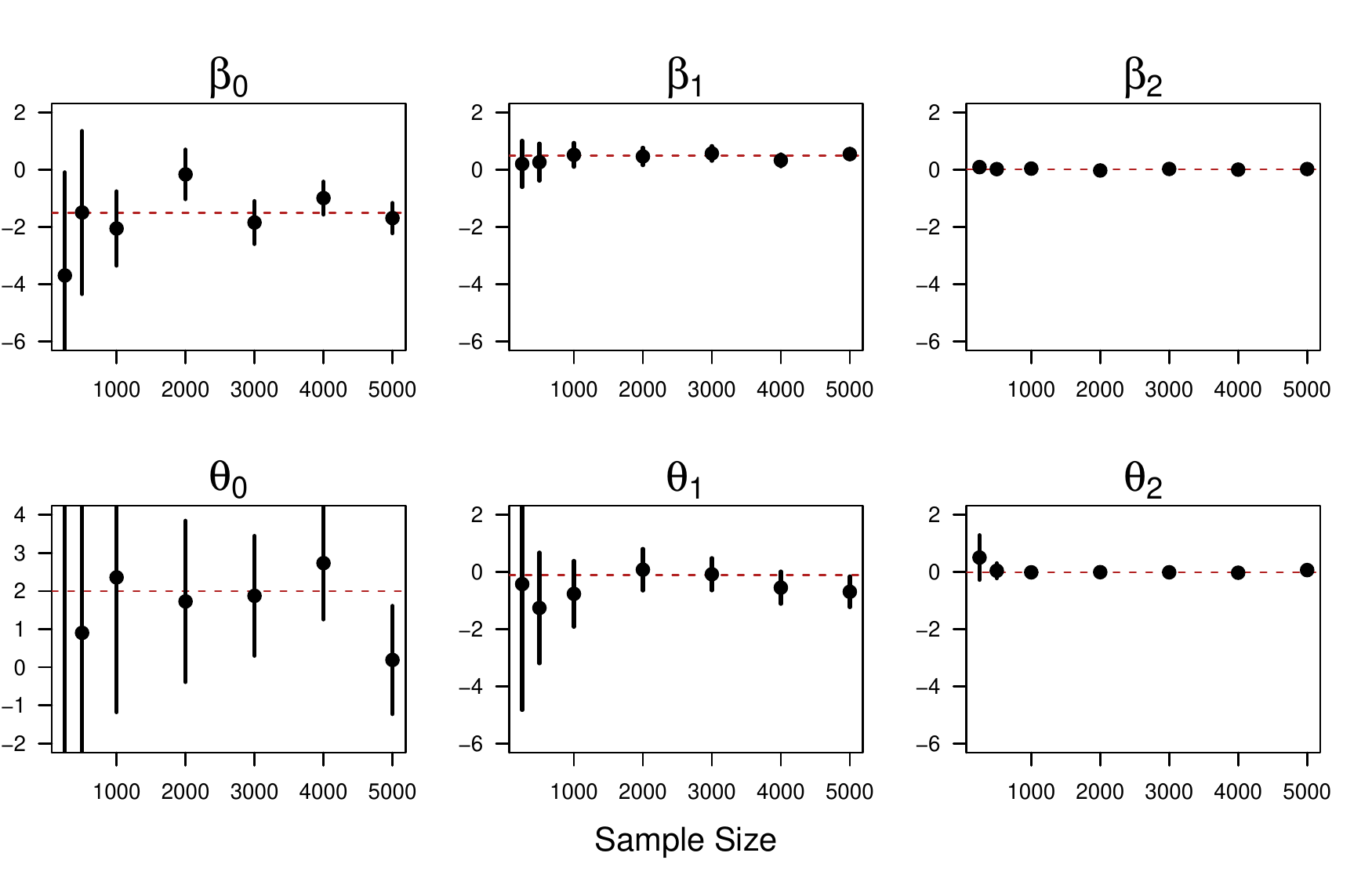}
\caption{ \textbf{Finite Sample Performance of Regression Estimator (Sensitive Trait as a Predictor).}    \textit{Note}: Regression estimates of six parameters in simulated data. The dashed lines indicate the true values for the parameters.}\label{fig:simulate_reg1}
\end{figure}

\begin{figure}[tbh!]
    \centering
    \includegraphics[width=17cm]{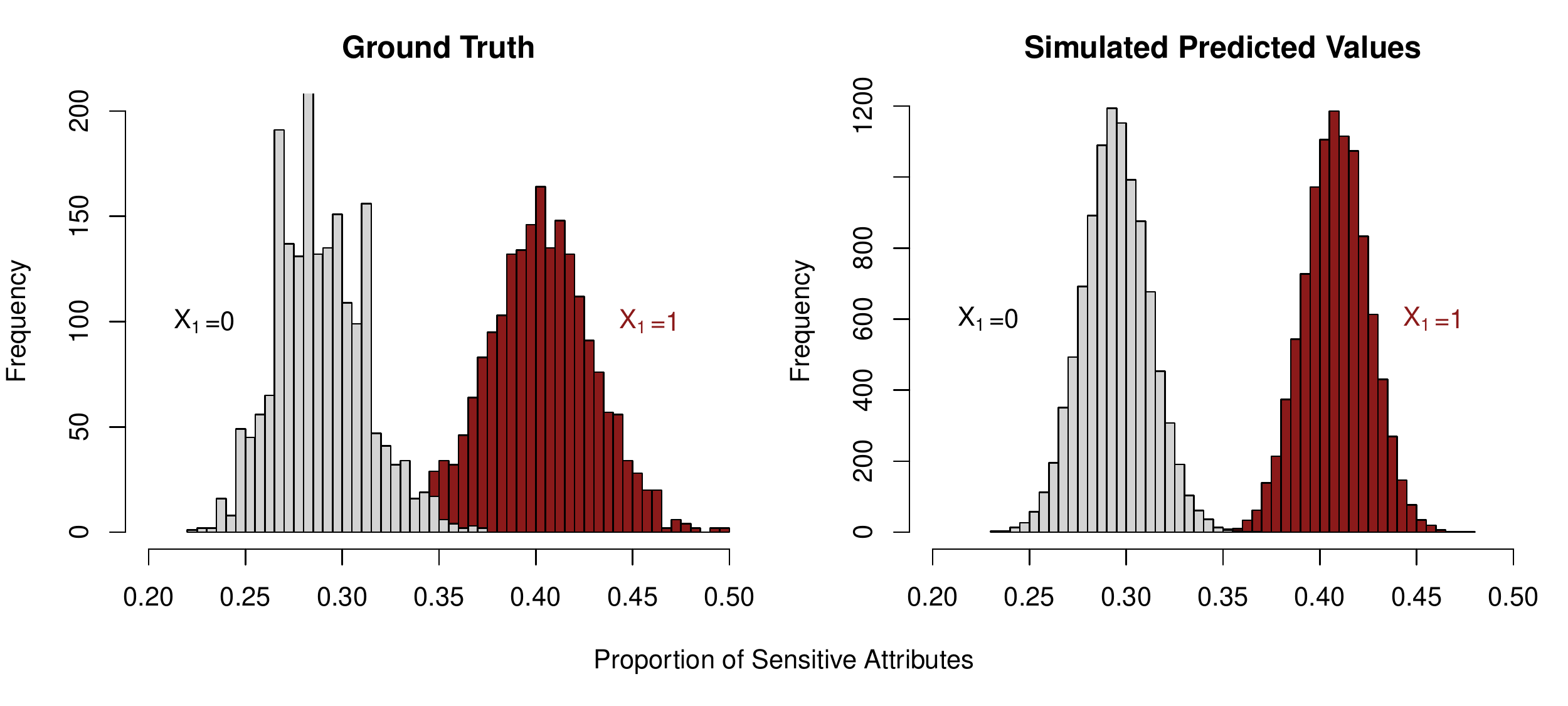}
\caption{ \textbf{True and Predicted Proportions of Sensitive Attributes}  \textit{Note}: This graph visualizes the empirical distribution of the probability of having sensitive attributes for $X_{1}=0$ and $X_{1}=1$ based on the ground truth with $N=4000$ (left panel) and simulated predicted values (right panel).}\label{fig:predicted1}
\end{figure}

Finally, we estimate the crosswise regression with the latent sensitive trait as the outcome variable. Figure \ref{fig:simulate_reg1} displays the estimated parameters and confidence intervals with different sample sizes. The results suggest that the proposed model and estimation strategy can recover the true parameters (asymptotically). It is also straightforward to compute predicted probabilities of having a sensitive attribute with 95\% confidence intervals using the parametric bootstrap. Figure \ref{fig:predicted1} displays the predicted probabilities of having sensitive attributes in this particular simulation with different values for $X_{1}.$\\

\subsection{Multivariate Regression: Sensitive Attribute as a Predictor}\label{app:regress_pred}

To validate the proposed framework, we simulate crosswise data with two covariates as in Online Appendix \ref{app:regress}. We then simulate the response variable according to the following generative model:
$$ V_{i} = \gamma_{0} + \gamma_{1}X_{1} + \gamma_{2}X_{2} + \delta Z_{i} + \epsilon_{i},$$

\noindent where we set $\gamma_{0}=0$, $\gamma_{1}=0.3$, $\gamma_{2}=0.01$, $\delta=1$, and $\epsilon_{i} \sim N(0,1)$. Recall that $Z_{i}$ is a latent variable for having a sensitive trait and we cannot observe its value directly (and thus crosswise data do not contain $Z_{i}$).

\begin{figure}[tbh!]
\centering
  
  \includegraphics[width=14cm]{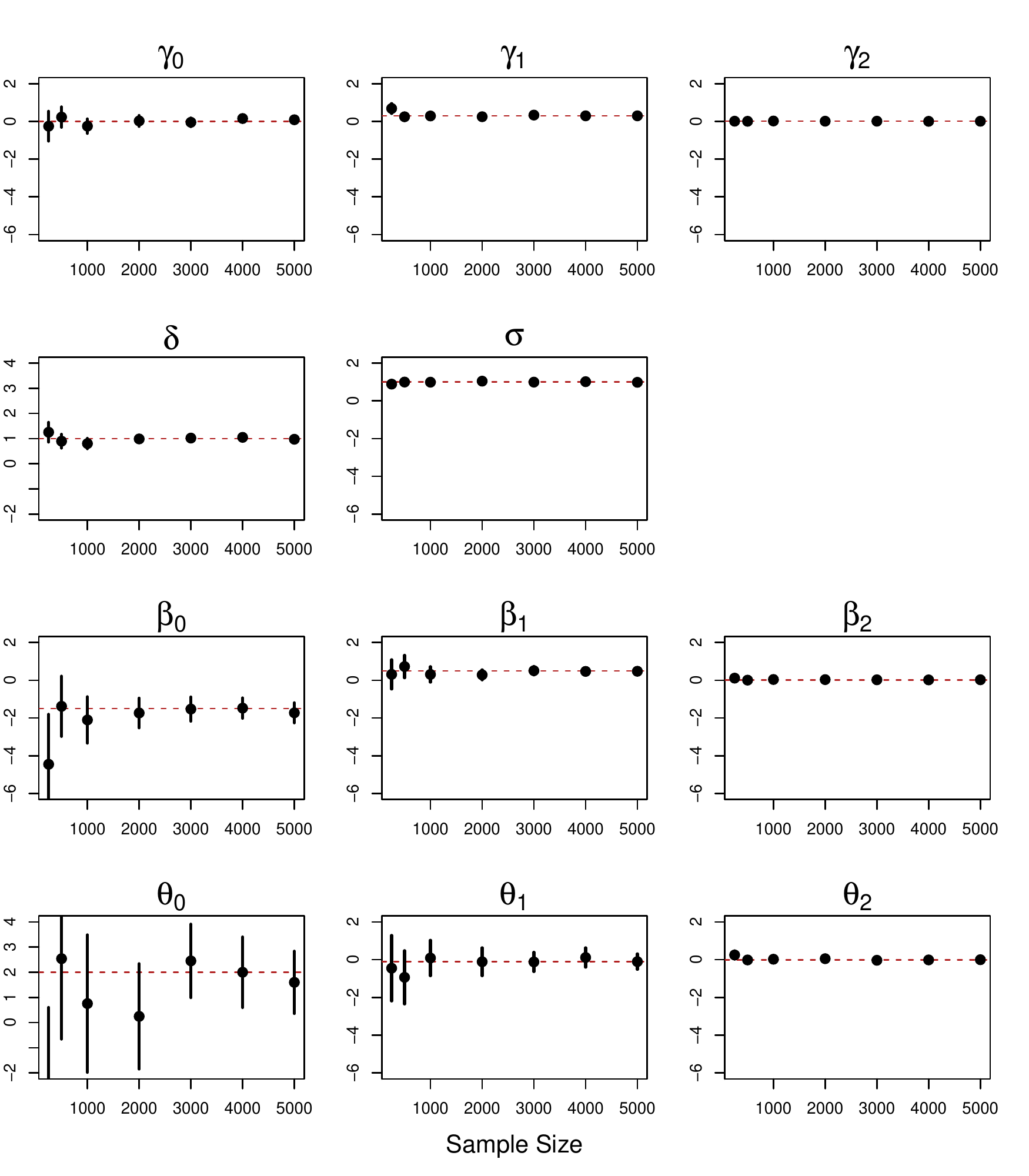}
\caption{ \textbf{Finite Sample Performance of Regression Estimator (Sensitive Trait as a Predictor)}. \textit{Note}: The dashed lines indicate the true values for the parameters.}\label{fig:simulate_reg2}
\end{figure}

\begin{figure}[tbh!]
    \centering
    \includegraphics[width=11cm]{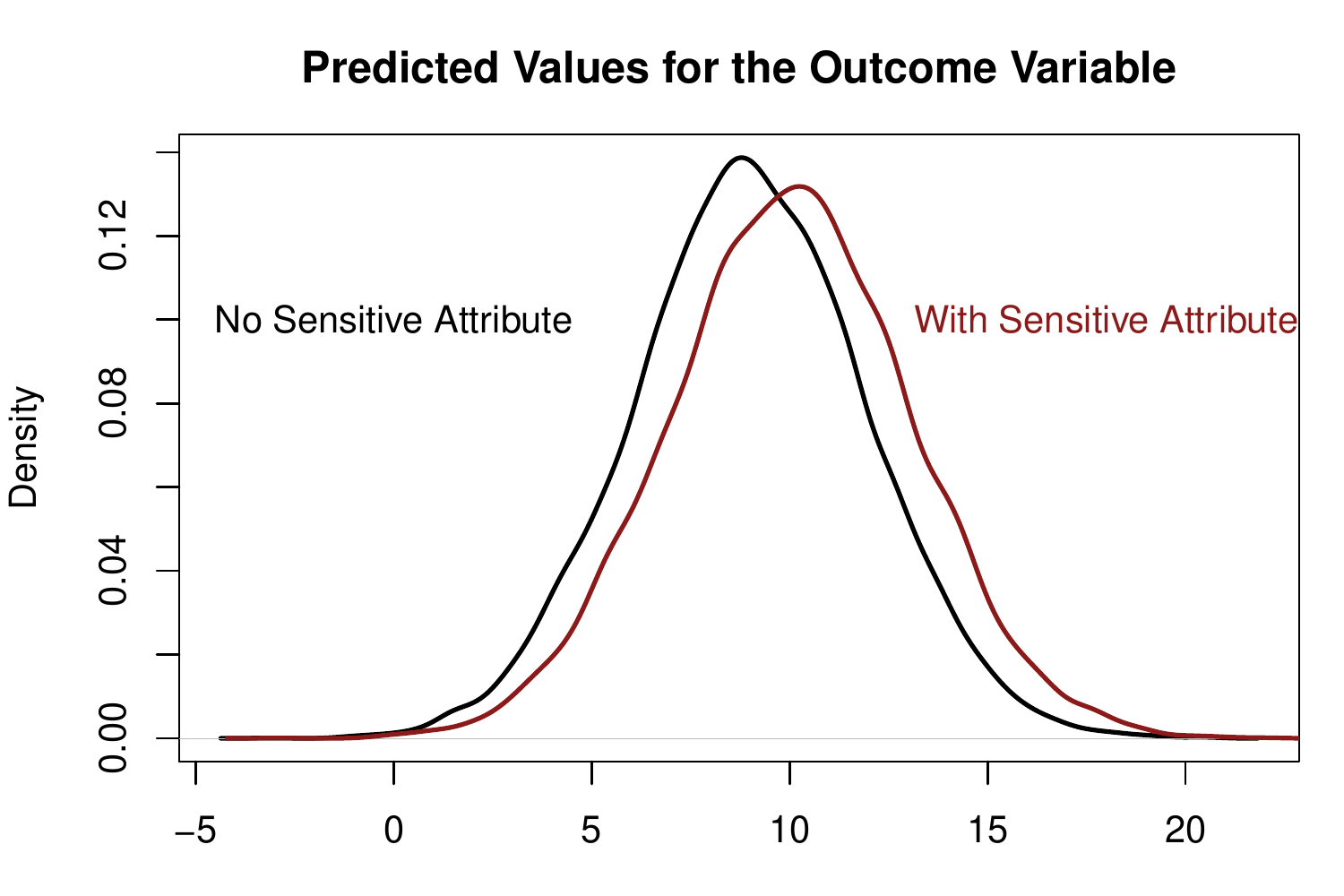}
\caption{ \textbf{Simulated Outcome Values with and without the Sensitive Attribute.}  \textit{Note}: This graph visualizes the density of simulated (predicted) values for the outcome variable in the absence (left density) and presence (right density) of the sensitive attribute.}\label{fig:predicted2}
\end{figure}

We then estimate the above crosswise regression model with the simulated observed outcome and crosswise data. Figure \ref{fig:simulate_reg2} shows the estimates for our quantity of interest with different sample size. It demonstrates that the proposed regression model and estimation strategy can recover the latent magnitude of the association between the latent sensitive trait and the response variable ($\delta=1)$. It also shows that other ten parameters can be properly estimated by the proposed regression model. It is also straightforward to compute predicted values of the outcome variable with 95\% confidence intervals using the parametric bootstrap. Figure \ref{fig:predicted2} displays the predicted values of the outcome in this particular simulation with different values for $Z$.  \\

\newpage
\subsection{Secret Number Approach to Non-Sensitive Statements}\label{app:secret}

This section introduces another extension of our method, which was not fully discussed in the main manuscript. One of the obstacles for some researchers to use the crosswise model is that they need to find an appropriate non-sensitive statement with a known prevalence in the crosswise question. Further, to maintain respondent privacy it is essential that a \textit{different} non-sensitive item be used with each crosswise question in the survey. This can be quite difficult to do in practice when one asks many crosswise questions.  To remedy this problem, we propose a \textit{secret number approach} to the non-sensitive statement. The essence of the secret number approach is to use a different virtual die roll for the non-sensitive crosswise item in each question, but to do so in a way that makes it clear to respondents the critical information remains private even if they believe surveyors are recording the result of the virtual die rolls.

If respondents believe that a virtual die roll is private, there are many advantages of using it in online surveys for the non-sensitive piece of private information needed in the crosswise question (e.g., after they roll the die the statement used is something like “the value of my die roll was 4” rather than a statement like “my mother was born in January.”) 
These advantages are the same ones as using a physical die in a paper and pencil survey and include:\\

\noindent $\bullet$	The necessary probabilities are known completely and do not change with the population being surveyed.\\

\noindent $\bullet$	The die roll can be used in the same format in many different questions, alleviating the need to develop a large library of items with known prevalence for surveys with many crosswise items.\\

\noindent $\bullet$	 There is no chance that respondents will not know the value of the die roll or be uncertain about it (unlike, for example, their mother's birthday).\\

\noindent $\bullet$	The die roll is independent of sensitive item by construction.\\

The problem with the virtual die roll is that respondents will often not believe that it is not being recorded and so their privacy protection will be undermined.

However, a simple addition to the instructions for rolling a virtual die alleviates this problem completely. To reassure respondents that such virtual dice rolls cannot be recorded and used by researchers to undermine privacy, one can simply ask the respondent to first pick (and perhaps write down) a ``secret number'', which must be an integer between 1 and 6, and then roll the virtual die. Then, the statement they evaluate is simply ``The value of the dice roll for this question was equal to my secret number.''

Further, in surveys with multiple crosswise items, respondents can be told that the can change their secret number on each question or not as they like. Since researchers will not know whether they have or not, their privacy is assured.

\clearpage
\section{Examples and Discussion of Constructing Anchor Questions}\label{sec:appAnchor}
\setcounter{figure}{0} 
\setcounter{table}{0}  
\setcounter{equation}{0} 
\setcounter{footnote}{0} 

As discussed in the main text, anchor questions should be topically connected to the other questions that are being asked in the crosswise format and should be sufficiently sensitive that they do not stand out relative to the other crosswise items.

Our purpose here is to first provide some examples of anchor questions that could be used for typical kinds of political science surveys that might use crosswise questions and then some discussion and practical advice about how to choose/construct good anchor questions. 

\indent

\subsection{Examples}

To help researchers get started on writing useful anchor questions, we provide a set of examples for anchor questions that might fit into crosswise survey modules on sensitive political topics. Of course, these examples are meant only as a starting point and may or may not make sense in any given setting. As always researchers should think carefully (and creatively) about the specific survey and sample in which they might be used.

\begin{figure}[thp!]
    \centering
    \includegraphics[width=18cm]{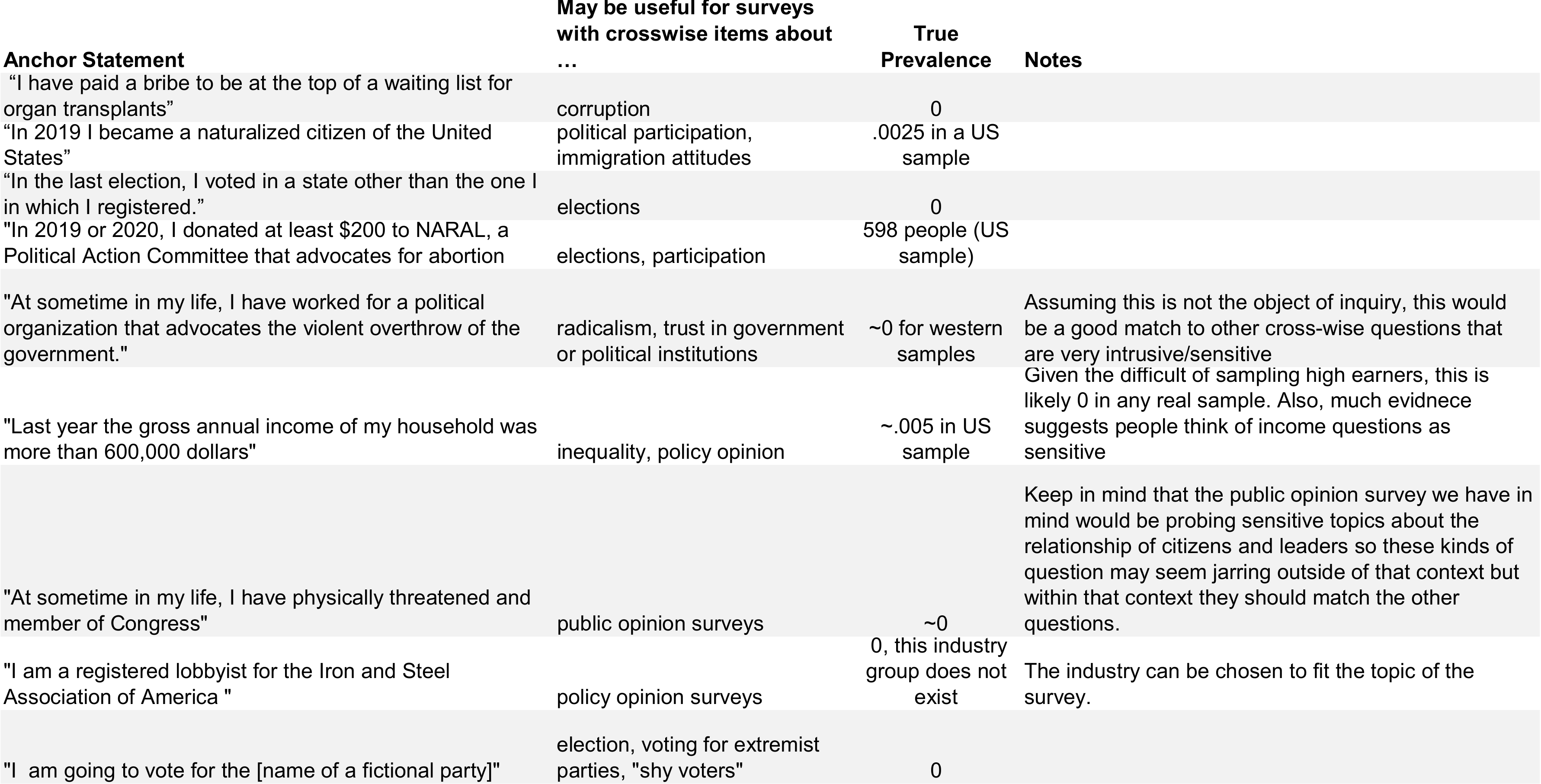}
\end{figure}

\indent

\subsection{Discussion}

The examples above illustrate some of the different kinds of sensitive statements that may have known prevalence in a given population.  As the diversity of the examples illustrate, however, statements differ in the extent to which they are sensitive and so they should be matched in terms of the level of sensitivity to the other crosswise questions in the survey. 

In addition, some of the examples suggest that researchers should try to avoid statements that, while they have might have a true prevalence of zero, might nevertheless attract false positives for reasons other than inattention. One way this can happen is if an item attracts support for purely  symbolic or affecting reasons. For example, in the example that uses a fictional party, care must be taken in choosing a neutral name of the fictional party. \citetSM{prior2015you}, for example, have shown that partisans sometimes report factual opinions on surveys that they know to be wrong as a kind of partisan cheerleading. Applying this to an election survey aimed at detecting ``shy'' Trump voters, for example, one might use an anchor question like ``In the 2020 election I voted for the Anarchy Party of Texas.''  Since there is no such party, the prevalence rate should be zero. However, we might expect this statement to attract some support from anarchists who engaged in a kind of partisan cheerleading. 

A better alternative would be a party with a more neutral name. This would be less sensitive, but in the context of a survey asking crosswise questions which other (extremist and non-extremist) parties one would vote for, it would likely fit in well. This illustrates again that whether an anchor question matches the set of crosswise questions well depends entirely on the context of the other crosswise questions.

Finally, it’s important to point out that a given anchor question can be made more typical of the whole set of crosswise questions not only by altering the anchor question itself but also by altering the composition of the whole set of crosswise questions in ways that make a given anchor question more representative of the whole set.

Specifically, those who are finding it difficult to construct a topically relevant anchor question that is sufficiently sensitive to match their crosswise questions of interest can simply add several topically relevant crosswise questions that are less sensitive than their crosswise questions of interest. In this way, the distribution of all the crosswise questions that the respondent sees (before encountering the anchor question) can be made to be more diverse in terms of sensitivity. This may make the anchor question less jarring when it is encountered. Likewise, if the trouble is finding a topically relevant anchor question, then several off-topic crosswise questions can be added to alter the mix of crosswise questions in ways that make the anchor question less of a stand-out.

\clearpage
\singlespacing
\bibliographystyleSM{apsr}
\bibliographySM{crosswise.bib}

\end{document}